\documentclass[11pt]{article}
\pdfoutput=1
\usepackage{amsmath,amssymb,color,epsfig,cite}
\usepackage{graphicx}
\usepackage{subfigure}
\usepackage{setspace}
\usepackage{amsthm,mathenv}
\usepackage{bbm}

\allowdisplaybreaks[4]

\usepackage{amsfonts}

\newcommand{\hoch}[1]{$\, ^{#1}$}

\makeatletter
\@addtoreset{equation}{section}
\makeatother

\newcommand{\be}{\begin{equation}}
\newcommand{\ee}{\end{equation}}
\newcommand{\bea}{\setlength\arraycolsep{2pt} \begin{eqnarray}}
\newcommand{\eea}{\end{eqnarray}}
\newcommand{\nn}{\nonumber}

\def\0{{\sst{(0)}}}
\def\1{{\sst{(1)}}}
\def\2{{\sst{(2)}}}
\def\3{{\sst{(3)}}}
\def\4{{\sst{(4)}}}
\def\5{{\sst{(5)}}}
\def\6{{\sst{(6)}}}
\def\7{{\sst{(7)}}}
\def\8{{\sst{(8)}}}
\def\sst#1{{\scriptscriptstyle #1}}

\begin{document}

\begin{center}
{\large {\bf Estimating the Final Spin of  Binary Black Holes Merger \\
 in STU Supergravity }}

\vspace{15pt}
{\large Shou-Long Li\hoch{1}, Wen-Di Tan\hoch{2},
Puxun Wu\hoch{1} and Hongwei Yu\hoch{1}}

\vspace{15pt}

\hoch{1}{\it Department of Physics and Synergetic Innovation Center for Quantum Effects and Applications, Hunan Normal University, Changsha, Hunan 410081, China }

\vspace{10pt}

\hoch{2}{\it Center for Joint Quantum Studies and Department of Physics,
School of Science, Tianjin University, Tianjin 300350, China }

\vspace{10pt}

\vspace{40pt}

\underline{ABSTRACT}

\end{center}

In this paper, we adopt the so-called Buonanno-Kidder-Lehner (BKL) recipe to estimate the final spin of a rotating binary black hole merger in STU supergravity.  According to the BKL recipe, the final spin can be viewed as the sum of the individual spins plus the orbital angular momentum of the binary system which could be approximated as the angular momentum of a test particle orbiting at the innermost stable circular orbit around the final black hole.  Unlike previous works, we consider the contribution of the orbital angular momentum of the binary system to the final spin by requiring the test particle to preserve the scaling symmetry in the Lagrangian of supergravity. We find some subtle differences between two cases corresponding to whether the symmetry is taken into account or not.
In the equal initial spin configuration, when the initial black holes  are non-spinning, the final spin of the merger is always larger than that in the case in which the symmetry is not imposed although the general behaviors are similar. The difference  increases firstly and then decreases as the initial mass ratio approaches unity. Besides, as the initial spins exceed a threshold, the final spin  is always smaller than that in the case where the scaling symmetry is not considered. The difference decreases constantly as the equal initial mass limit is approached. All these features exist in the merger   of a binary STU black hole with different charge configurations. We also study the final spin's difference between different charge configurations and different initial spin configurations.

\vfill
 shoulongli@hunnu.edu.cn\ \ \ TWDHANNAH@163.com \ \ \ pxwu@hunnu.edu.cn \ \ \ hwyu@hunnu.edu.cn

\thispagestyle{empty}

\pagebreak



\newpage

\section{Introduction}

The evolution of the coalescence of a binary black hole (BBH) system is widely accepted to include three stages: inspiral, merger and ringdown.
The early inspiral and ringdown stages can be well explained by the post-Newtonian approximation~\cite{Blanchet:2013haa} and the black hole perturbation theory~\cite{Regge:1957td, Zerilli:1971wd, Teukolsky:1973ha, Teukolsky:1974yv}, respectively.
Since the late inspiral and merger stages are highly nonlinear, only numerical relativity simulations could provide an accurate description of the dynamical properties of the whole process~\cite{Lehner:2014asa}.
The numerical simulations have been widely used in the study of the discovered gravitational-wave (GW) events, such as those in Refs.~\cite{Abbott:2016blz,  Abbott:2016nmj, Abbott:2017vtc, Abbott:2017gyy, Abbott:2017oio}.
 However,  full simulations have been known to be highly-costly, and take a lot of time.
 This inspires one to look for some reliable though may-not-so-rigorous methods to reproduce reasonably accurate results compared with those from available numerical simulations. One expects that such a method  not only  can give some useful predictions for the final state, but also is  helpful to providing an accurate analytic template.  In this regard, the Buonanno-Kidder-Lehner (BKL) recipe~\cite{Buonanno:2007sv} provides a simple first-principles-derived method to estimate the final spin of the merger which is one of the most important  properties of the remnant black hole that could help detection~\cite{Barausse:2009uz, Berti:2007zu} and distinguish the BBH from  other exotic objects\cite{Krishnendu:2017shb}. The advantage of  the recipe is that it can be applied to the merger of a BBH with arbitrary initial masses and spins.  Based on the approximate conservation of mass and angular momentum of a BBH system during the merger and ringdown phases, and  some other simple assumptions, the BKL recipe can be used to  straightforwardly and accurately estimate the final angular momentum as a sum of the individual spins plus the orbital angular momentum of the binary system which could be approximated as the angular momentum of a test particle orbiting at the innermost stable circular orbit (ISCO) around the final rotating black hole.  This point-particle approximation captures the key aspects of two-body dynamics, and the method is also supported by the numerical simulations~\cite{Buonanno:2000ef, Damour:2007cb, Buonanno:2007pf, Buonanno:2006ui, Davis:1971gg, Davis:1972ud, Price:1994pm, Baker:1996bt, Baker:2001nu}.

  By considering that the test particle is charged, the BKL recipe has been generalized to estimate the final spin of the binary charged black hole merger in the Einstein-Maxwell (EM) theory~\cite{Jai-akson:2017ldo}, as well as in the  Kaluza-Klein~(KK) theory~\cite{Jai-akson:2017ldo} and the low energy limit  of the heterotic string theory~\cite{Siahaan:2019oik}.  With the BKL recipe, the estimates for the final spin of a binary black hole merger in different modified gravities are expected to be different, and thus could be constrained by the observations. Although recent observations have been found to support general relativity, some subtle deviations may be probed as the higher signal-to-noise ratio will be achieved  in the near future. So, the final spin of a BBH merger may  provide a possible way to test the string theory and other modified gravities near strong gravitational regimes.

In this work, we would like to revisit the details of the generalized BKL recipe to estimate the final spins of a binary charged black hole merger.  In the previous work~\cite{Jai-akson:2017ldo},  the Lagrangian of the charged test particle  in the KK theory is taken  the  same as that in EM theory. However, it is worth noting that  symmetry plays a very important role when we study the black hole in the framework of string theory and supergravities~\cite{Youm:1997hw, Mohaupt:2000gc, Peet:2000hn}.  For example, after performing a dimensional reduction~\cite{Lu:1998xt, Cremmer:1999du} on $S^1$ from five-dimensional pure Einstein gravity, the resulting four-dimensional Einstein-Maxwell-dilaton theory with a special coupling constant, i.e. the KK theory, has an extra scaling symmetry, namely a constant shift of the dilaton accompanied by an appropriate constant scaling of the Maxwell potential. This symmetry can be understood in terms of {\color{red} $\mathbb R^n (n =1)$~\cite{Chow:2014cca}.} If we further reduce the four-dimensional KK theory to the three-dimensional theory, the corresponding global symmetry  $SL(3, R)$ can help us to explore solutions of the KK theory. So, it is worth examining the outcome from the BKL recipe  when the Lagrangian describing the motion of the test particle  also preserves the same symmetry~\cite{Lu:2019icm}.  In this case, the angular momentum of the test particle may be modified. It is natural to ask whether the revised method could improve the precision of the estimation  by comparing with the numerical simulations~\cite{Hirschmann:2017psw, Hofmann:2016yih}. As a first step, we will study the difference of the final spin estimations between the two cases, i.e, the case  where  the scaling symmetry is taken into account and the one which is not.

On the other hand, it is worth noting that compared with  Kerr-Newman (KN) black hole in EM theory, black holes in supergravities and string theory have extra scalar charges. When we study the binary dynamics in supergravities and string theory, it is necessary to consider the effects of scalar fields~\cite{Lu:2019icm} apart from those of electromagnetic fields and gravitational fields. This could be achieved by writing the Lagrangian of test particle in an appropriate metric frame.

The four-dimensional EM theory, the KK theory and the low energy limit of the heterotic string theory can be viewed as the special cases of a more general supergravity, i.e. STU supergravity~\cite{Chong:2004na, Duff:1995sm, Cvetic:1996kv, Chow:2014cca},  which has an $SL(2, R)^3$ symmetry and also the so-called ``S-T-U'' triality symmetry under permutations of the three $SL(2, R)$ factors, and can be obtained from higher dimensional string theory and carries four independent electromagnetic fields. To be specific,  STU rotating black holes carrying four equal charges, two equal charges and a single charge are equivalent to the KN black hole, the Einstein-Maxwell-dilaton-axion (EMDA) black hole (which is equivalent to Kerr-Sen black hole in the low energy limit of the heterotic string theory)~\cite{Sen:1992ua} and the KK black hole~\cite{Rasheed:1995zv} respectively. Here we will estimate the final spin of the binary STU black hole merger by using the BKL recipe and requiring the Lagrangian of the test particle to preserve the scaling symmetry in STU supergravity.  Let us note here that a binary charged black hole could possibly be the source to produce the counterpart  electromagnetic signal to the merger of BBH, which could be used to explain the signal recently observed by the Fermi Gamma-ray Burst Monitor (GBM) group~\cite{Connaughton:2016umz, Loeb:2016fzn, Perna:2016jqh, Zhang:2016rli, Woosley:2016nnw, Liu:2016olx}. So it is meaningful to explore the final spin of the merger of a binary charged black hole, at least BBH with weak charges ($Q\ll M$).

The organization of the paper is  as follows. In section~\ref{sec2}, we review the rotating black hole in the four-dimensional STU supergravity. In section~\ref{sec3}, we review the BKL recipe and reconsider the contribution of the orbital angular momentum of the binary system to the final spin by requiring the test particle  to preserve the scaling symmetry in STU supergravities. In section~\ref{sec4}, we first study the ISCO of the test particle, and then estimate the final spin of binary STU rotating black holes with different charge configurations in different initial spin cases such as equal initial spins, unequal initial spins and generic initial spins. We conclude in section~\ref{sec5}.

\section{STU supergravity} \label{sec2}

The four-dimensional Lagrangian for the bosonic sector of the ${\cal N}=2$ supergravity coupled to three vector multiplets, also called  the STU model,  is given by~\cite{Chong:2004na, Cvetic:1996kv, Chow:2014cca}
\bea
 { L}_{\textup{STU}} &=& R \star {\mathbbm 1} - \frac12 \sum_{i=1}^3 (\star d \varphi_i \wedge d \varphi_i  + e^{2 \varphi_i} \star d \psi_i \wedge d \psi_i )  -\frac{1}{2}  e^{-\varphi_1 }  \big( e^{ \varphi_2 -\varphi_3 } \star \hat{F}_1 \wedge \hat{F}_1 \nn  \\
 && +e^{ \varphi_2 +\varphi_3 } \star \hat{F}_2 \wedge \hat{F}_2 +e^{ -\varphi_2 +\varphi_3 } \star \hat{\cal F}^1 \wedge \hat{\cal F}^1 +e^{ -\varphi_2 -\varphi_3 } \star \hat{\cal F}^2 \wedge \hat{\cal F}^2 \big)  \nn \\
 && -\psi_1 \big( \hat{F}_1 \wedge \hat{\cal F}^1 +  \hat{F}_2 \wedge \hat{\cal F}^2 \big) \,, \label{stu}
\eea
where $\varphi_i$ and $ \psi_i$ are dilatons and axions respectively. The four  field strengths can be written in terms of potentials as
\bea
\hat{F}_1 &=& d \hat{A}_1 - \psi_2 d\hat{\cal A}^2  \,, \quad \hat{F}_2 = d \hat{A}_2 + \psi_2 d\hat{\cal A}^1 -  \psi_3 d\hat{ A}_1  + \psi_2 \psi_3 d\hat{\cal A}^2  \,, \nn \\
\hat{\cal F}^1 &=& d\hat{\cal A}^1 + \psi_3 d\hat{\cal A}^2 \,, \quad \hat{\cal F}^2 = d\hat{\cal A}^2 \,.
\eea
The rotating STU black hole solution is given by~\cite{Chong:2004na}
\bea
&& ds^2 = -\frac{\rho^2 - 2 m r}{W}(d t+{\cal B} )^2 +W \left(\frac{d r^2}{\Delta} +d\theta^2 +\frac{\Delta \sin^2 \theta d\phi^2}{\rho^2 - 2 m r} \right)  \,, \nn \\
&& \hat{A}_1 = A_1 +\sigma_1 {\cal B}  +\sigma_1 d t \,, \quad  \hat{A}_2 = A_2 +\sigma_2 {\cal B}  +\sigma_2 d t \,,  \nn \\
&& \hat{\cal A}^1 = {\cal A}^1 +\sigma_3 {\cal B}  +\sigma_3 d t \,, \quad  \hat{\cal A}^2 = {\cal A}^2 +\sigma_4 {\cal B}  +\sigma_4 d t \,,  \nn \\
&& \psi_1 = \frac{2 m u (c_{13} s_{24} -c_{24}s_{13})}{r_1 r_3 +u^2} \,, \quad  \psi_2 = \frac{2 m u (c_{14} s_{23} -c_{23}s_{14})}{r_2 r_3 +u^2} \,, \quad  \psi_3 = \frac{2 m u (c_{12} s_{34} -c_{34}s_{12})}{r_1 r_2 +u^2} \,, \nn \\
&& \varphi_1 = \ln \frac{r_1 r_3 +u^2}{W} \,, \quad  \varphi_2 = \ln \frac{r_2 r_3 +u^2}{W} \,, \quad  \varphi_3 = \ln \frac{r_1 r_2 +u^2}{W} \,,
\eea
where
\bea
&&{\cal B} = \frac{2 m (a^2 -u^2) (r c_{1234}- (r -2 m) s_{1234} ) }{ a (\rho^2 -2 m r) } d\phi \,, \nn \\
&&A_1 = -\frac{2 m u c_1 s_1 \Delta d\phi}{ a (\rho^2 - 2 m r) }  \,,\quad  A_2 = -\frac{2 m u (a^2 - u^2) ( (r- 2 m ) c_2 s_{134} -r c_{134} s_2) d\phi }{ a (\rho^2 - 2 m r) }  \,, \nn \\
&&{\cal A}^1 = -\frac{2 m u c_3 s_3 \Delta d\phi}{ a (\rho^2 - 2 m r) }   \,,\quad  {\cal A}^2 = -\frac{2 m u (a^2 - u^2) ( (r- 2 m ) c_4 s_{123} -r c_{123} s_4) d\phi }{ a (\rho^2 - 2 m r) }  \,, \nn  \\
&&\sigma_1 = \frac{ 2 m u }{ W^2} \left( (r r_1 +u^2) (c_{234} s_1 -s_{234} c_1 ) +2 m r_1 s_{234} c_1 \right) \,, \nn \\
&&\sigma_2 = \frac{ 1 }{ W^2} \left(2 m c_2 s_2 (r_1 r_3 r_4 +r u^2) +4 m^2 u^2 e_2 \right) \,, \nn \\
&&\sigma_3 = \frac{ 2 m u }{ W^2} \left( (r r_3 +u^2) (c_{124} s_3 -s_{124} c_3 ) +2 m r_3 s_{124} c_3 \right) \,, \nn \\
&&\sigma_4 = \frac{ 1 }{ W^2} \left(2 m c_4 s_4 (r_1 r_2 r_3 +r u^2) +4 m^2 u^2 e_4 \right) \,, \nn \\
&&e_2 = c_{134} s_{134} (c_2^2 +s_2^2) -c_2 s_2 (s_{13}^2 +s_{14}^2 +s_{34}^2 +s_{134}^2  ) \,, \nn \\
&&e_4 = c_{123} s_{123} (c_4^2 +s_4^2) -c_4 s_4 (s_{12}^2 +s_{13}^2 +s_{23}^2 +s_{123}^2  ) \,, \nn \\
&&W^2 = \prod_{i=1}^4 r_i +u^4 +2 u^2 \left(  r^2 +  m r \sum_{i=1}^4 s_i^2  + 4 m^2 (\prod_{i=1}^4 c_i s_i -\prod_{i=1}^4 s_i^2) - 2 m^2  \sum_{i<j<k}^4 s_{i}^2 s_{j}^2 s_{k}^2 \right) \,, \nn \\
&&\Delta = r^2 - 2 m r + a^2 \,, \quad \rho^2 = r^2 +a^2 \cos^2 \theta \,, \quad r_i = r +2 m s_i^2 \,, \quad u = a \cos \theta \,, \nn \\
&&c_{1\dots n} = \cosh \delta_1 \dots \cosh \delta_n  \,, \quad s_{1\dots n} = \sinh \delta_1 \dots \sinh \delta_n  \,,
\eea
where parameters $(m, a, \delta_1, \delta_2, \delta_3, \delta_4)$ characterize mass, angular momentum, and four electric charges respectively.

The four-dimensional theory~(\ref{stu}), which can be obtained from the one in six-dimensions by reducing the bosonic string on $T^2$,  has a global symmetry  $ SL(2, \mathbbm R)^3$, realized nonlinearly on the scalar coset $ (SL(2, \mathbbm R)/U(1))^3$~\cite{Chong:2004na}.  It is worth noticing that the local general coordinate symmetry in the six-dimensional bosonic string involves coordinate reparameterisations by arbitrary functions of six coordinates, while the local general coordinate symmetry and $U(1)$ gauge transformations in four dimensions  involve arbitrary functions of only four coordinates. Actually the theory has another symmetry, namely  a constant shift of the dilaton fields $\varphi_i$, accompanied by an appropriate constant scaling of  the axion fields $\psi_i$ and the Maxwell potentials:
\bea
&&\varphi_1 \rightarrow \varphi_1 +\alpha_1 \,,\quad \varphi_2 \rightarrow \varphi_2 +\alpha_2 \,,\quad \varphi_3 \rightarrow \varphi_3 +\alpha_3 \,, \nn \\
&&\psi_1 \rightarrow \psi_1 e^{-\alpha_1} \,,\quad  \psi_2 \rightarrow \psi_2 e^{-\alpha_2} \,,\quad  \psi_1 \rightarrow \psi_3 e^{-\alpha_3} \,, \nn \\
&&\hat{A}_1 \rightarrow  \hat{A}_1 e^{\frac{\alpha_1 - \alpha_2 +\alpha_3}{2}} \,,\quad \hat{A}_2 \rightarrow  \hat{A}_2 e^{\frac{\alpha_1 - \alpha_2 -\alpha_3}{2}} \,, \nn \\
&& \hat{\cal A}^1 \rightarrow  \hat{\cal A}^1 e^{\frac{\alpha_1+\alpha_2 -\alpha_3}{2}} \,,\quad \hat{\cal A}^2 \rightarrow  \hat{\cal A}^2 e^{\frac{\alpha_1 +\alpha_2 +\alpha_3}{2}} \,,
\eea
which is important to studying the supergravity and the solutions of the theory. For our purpose to study the  binary STU black hole merger, we  would also like to consider the effect of the symmetry on the final spin estimation of  the remnant black hole.
For simplicity, we focus on  some special cases of the binary STU rotating black holes with different charge configurations. First, we consider the STU rotating black hole with a single non-zero charge, i.e. $\delta_4 =\delta \ne 0 \ (\sinh \delta = s, \cosh \delta = c), \delta_1 =\delta_2 =\delta_3 =0$. The solution reduces to the KK rotating black hole~(the full solution is given in the Appendix~\ref{app1}), and the corresponding theory is described by
\be
{\cal L}_{\textup{KK}} =  R -\frac12 (\nabla\varphi)^2 - \frac14 e^{-\sqrt{3} \varphi} F^2 \,,
\ee
where the canonically-normalized electromagnetic field $F = {\cal F}^2, \varphi = \sqrt{3} \varphi_1 =\sqrt{3} \varphi_2 =\sqrt{3} \varphi_3, \psi_1 =\psi_2 =\psi_3 =0,  {\cal F}^1 = {F}_1  = {F}_2 =0$, and we have for the physical mass $M$, charge $Q$, and angular momentum $J$,
\be
M = \frac{m}{2}(2+s^2) \,,\quad Q = \frac{ m s c}{2} \,,\quad J = m a  c \,.
\ee
The corresponding scaling symmetry is given by
\be
\varphi \rightarrow \varphi + \alpha \,,\quad A \rightarrow A e^{\frac{\sqrt{3}}{2} \alpha} \,.
\ee
Second, we consider the STU rotating black hole with two non-zero equal charges, i.e. $\delta_2 =\delta_4 =\delta \ne 0, \delta_1 =\delta_3  =0$. The solution reduces to the EMDA rotating black hole~(the full solution is given in the Appendix~\ref{app1}), and the corresponding theory is governed by
\be
{\cal L}_{\textup{EMDA}} =  
R -\frac12 (\nabla\varphi)^2 -\frac12 e^{2 \varphi} (\nabla\psi)^2 -   \frac14 e^{- \varphi} F^2 - \frac18 \psi \epsilon^{\mu\nu\rho\sigma} F_{\mu\nu} F_{\rho\sigma} \,,
\ee
where $ \epsilon^{\mu\nu\rho\sigma}$ is the Levi-Civita tensor, the canonically-normalized electromagnetic field $F = {\cal F}^2 /\sqrt{2} = {F}_2 /\sqrt{2}, \varphi =  \varphi_1, \psi = \psi_1 , {\cal F}^1 = {F}_1   =0 , \varphi_2 =\varphi_3 =0, \psi_2 =\psi_3 =0,$ and we have
\be
M = m c^2 \,, \quad Q = \frac{\sqrt{2}}{2} m s c \,, \quad J = m c^2 a \,.
\ee
The corresponding scaling symmetry reduces to
\be
\varphi \rightarrow \varphi + \alpha \,,\quad A \rightarrow A e^{\frac{1}{2} \alpha} \,, \quad  \psi \rightarrow \psi e^{ - \alpha}  \,.
\ee
Third, we consider the STU rotating black hole with four non-zero equal charges, i.e. $\delta_1 =\delta_2 =\delta_3  =\delta_4 =\delta \ne 0$, the solution reduces to the KN black hole~(the full solution is given in the Appendix~\ref{app1}) after making a coordinate transformation $r \rightarrow r+ 2 m s^2$ and applying the electromagnetic duality, and the corresponding theory is described by
\be
{\cal L}_{\textup{EM}} =   R  - \frac{1}{4} F^2 \,,
\ee
where the canonically-normalized electromagnetic field $F = {\cal F}^1 = {F}_1  = {\cal F}^2 = {F}_2  , \varphi_1 = \varphi_2 =\varphi_3 =0, \psi_1 =\psi_2 =\psi_3 =0,$ and we have
\be
M = m  (s^2 +c^2)  \,, \quad  Q=  m s c \,, \quad J = m  (s^2 +c^2) a \,.
\ee
There is no similar scaling symmetry in the EM theory because all of the dilatons and axions vanish.

\section{BKL recipe in STU supergravity} \label{sec3}

In this section we will first review the BKL recipe  and consider the contribution of the orbital angular momentum of the binary system to the final spin by requiring the Lagrangian of the test particle to preserve the scaling symmetry mentioned before. Then we will examine the Newtonian limit of the motion of the test particle orbiting the final black hole.

The BKL recipe~\cite{Buonanno:2007sv} was proposed to estimate  the final spin of a BBH merger with arbitrary initial masses and spins based on first principles and a few safe assumptions. One assumes that the BBH system evolves quasi-adiabatically, and radiates much angular momentum, which causes the binary orbit to become smaller gradually during the inspiral stage until it reaches the ISCO. Once the ISCO radius is reached, the binary orbit becomes unstable and a  ``plunge'' occurs, resulting in the BBH merger, and then the final black hole forms quickly. The loss of mass and angular momentum with respect to the total mass and angular momentum  of the binary system is small during the merger stage, and so it is reasonable to argue that mass and angular momentum are conserved approximately. One can also assume that the magnitude of the individual spins of the black holes remains constant because both spin-spin and spin-orbit couplings are small, and the radiation falling into the black holes affects the spins by a small amount.  Therefore, the mass $M$ of the final black hole can be given by
\be
M = M_1 +M_2 \,,
\ee
where $M_1$ and $M_2$ are the masses of initial black holes. The final mass $M$ can also be  described by a more accurate expression~\cite{Kesden:2008ga}. As mentioned before, the loss of the mass is small during the whole stages and this is a  good approximation to the first order in the gravitational wave observations~\cite{Abbott:2016blz, Abbott:2016nmj, Abbott:2017vtc, Abbott:2017gyy, Abbott:2017oio}. Moreover, the contribution of the orbital angular momentum to the final angular momentum of the black hole remnant can be described by the angular momentum of a test particle orbiting around the final rotating black hole at ISCO. The conservation of angular momentum at the moment of  plunge implies that
\be
M {\cal A}_f  = L_{\textup{orb}} +M_1  {\cal A}_1 +M_2  {\cal A}_2  \,,
\ee
where ${\cal A}_f $ is the spin of the final black hole, ${\cal A}_1 $ and $ {\cal A}_2$ are spins of initial black holes, and $L_{\textup{orb}}$ is the orbital angular momentum of the binary system which is represented by the angular momentum of a test particle with reduced mass $\mu = M_1 M_2/M$ orbiting around the final black hole at ISCO.
The final spin can be written as
\be
{\cal A}_f  = {\cal L} \nu +\frac{M \chi_1}{4} (1+\sqrt{1-4 \nu})^2 +\frac{M \chi_2}{4} (1-\sqrt{1-4 \nu})^2 \,, \label{finalspin}
\ee
where  ${\cal L} =L_{\textup{orb}}/\mu$ is the angular momentum of the test particle with unit mass,  $\chi_i = A_i /M_i (i= 1, 2) $, and $\nu = \mu /M $.

The contribution of the orbital angular momentum of the binary charged black hole  to the final spin can be effectively described by the angular momentum of a charged test particle orbiting around the final black hole~\cite{Jai-akson:2017ldo}. Furthermore, it is worth considering that the charged test particle preserves the symmetry in the Lagrangian of gravities when we discuss the particle motion in the framework of supergravity~\cite{Lu:2019icm}. Once the symmetry is taken into account, the particle motion and the orbital angular momentum are expected to be modified. Now we examine the Newtonian limit  and  identify reasonable mass and
charge assignments for our consideration.

Generically, the motion of a relativistic particle of mass $\mu$ coupled to the Maxwell field $A$ with charge $q$ is governed by the action
\be
{\cal S}_0 = \int  d\tau  \left( -\mu \sqrt{ -   g_{\lambda\nu} \dot{X}^\lambda \dot{X}^\nu } -\frac14 q A_\nu \dot{X}^\nu \right)   \,,
\ee
where $\tau$ and $X$ represent the proper time and coordinate respectively, and the dot denotes the derivative with respect to $\tau$. For our purpose to consider the test particle  that preserves the symmetry, the action should be given by~\cite{Lu:2019icm}
\be
{\cal S}_1 = \int  L_1 d\tau   = \int  d\tau  \left( -\mu \sqrt{ -  e^{\beta \varphi} g_{\lambda\nu} \dot{X}^\lambda \dot{X}^\nu } -\frac14 q A_\nu \dot{X}^\nu \right)   \,.
\ee
Compared with ${\cal S}_0$, the action ${\cal S}_1$ has the symmetry
\be
\varphi \rightarrow \varphi + \alpha \,,\quad A_\nu \rightarrow A_\nu e^{\frac{1}{2} \alpha\beta } \,. \label{homothety}
\ee
It is worth noting that this type of symmetry is akin to a homothety. Under the homothetic transformation~(\ref{homothety}), the action ${\cal S}_1$ will have an overall factor $e^{\alpha\beta/2 }$. In order to match the symmetry in previous cases, the values of $\beta $ are
\be
\textup{KK}: \beta = \sqrt{3} \,,\quad \textup{EMDA}: \beta = 1 \,,\quad \textup{EM}: \beta = 0  \,.
\ee
The related geodesic equation becomes
\be
 \mu \left( \ddot{X}^\mu  +{\Gamma}^\mu {}_{\rho\sigma} \dot{X}^\rho \dot{X}^\sigma \right) -\frac{\mu\beta}{2} (\partial^\mu\varphi +\partial_\lambda \varphi \dot{X}^\lambda \dot{X}^\mu) = \frac14 q e^{-\frac{1}{2} \alpha\beta } \dot{X}^\nu {F}_\nu {}^\mu
\ee
 which is invariant under the transformation~(\ref{homothety}).  
Note that the action ${\cal S}_1$ is difficult to quantize because it contains a square root, and cannot be used to describe a massless particle. Classically, this action ${\cal S}_1$ is equivalent to
\be
{\cal S}_2  = \int  L_2 d\tau  =  \int  d\tau   \left( \frac12 \xi^{-1}  e^{\beta \varphi} g_{\lambda\nu} \dot{X}^\lambda \dot{X}^\nu -\frac12 \mu^2 \xi  -\frac14 q A_\nu \dot{X}^\nu  \right)  \,.
 \ee
where  $\xi(\tau)$ is  the auxiliary field.  The homothetic symmetry for  action ${\cal S}_2$ becomes
\be
\varphi \rightarrow \varphi + \alpha \,,\quad A_\nu \rightarrow A_\nu e^{\frac{1}{2} \alpha\beta }  \,,\quad \xi \rightarrow \xi e^{\frac{1}{2} \alpha\beta } \,. \label{homothety2}
\ee
 We  can introduce a new metric
\be
\tilde{g}_{\lambda\nu} = e^{\beta \varphi} g_{\lambda\nu}  \,,
\ee
where the conformal factor $e^{\beta \varphi}$ takes into account precisely the effect of the dilaton. The weights of $(\tilde{g}_{\lambda\nu}, \tilde{g}^{\lambda\nu}, A_\nu, A^\nu, \xi)$ are $(1, -1, 1/2, -1/2, 1/2)$, respectively. Here $\tilde{g}^{\lambda\nu}$ and  $\tilde{g}_{\lambda\nu}$ have opposite weights, and we will let  $\tilde{g}^{\lambda\nu}$ and  $\tilde{g}_{\lambda\nu}$ raise and lower the indexes in the remaining discussions.
By varying the action ${\cal S}_2$ with respect to the auxiliary field $\xi$, the corresponding equation of motion is found to be
\be
\xi^2 \mu^2 +  \tilde{g}_{\lambda\nu} \dot{X}^\lambda \dot{X}^\nu =0 \,.\label{xigauge}
\ee
Solving the above equation for $\xi$ and substituting the solution back into ${\cal S}_2$ gives the original action ${\cal S}_1$. Varying both actions ${\cal S}_1$ and ${\cal S}_2$ with respect to $X^\mu$  yields the same equation of motion
\be
\mu \left( \ddot{X}^\lambda  +\tilde{\Gamma}^\lambda {}_{\rho\sigma} \dot{X}^\rho \dot{X}^\sigma \right) = \frac14 q \dot{X}^\nu \tilde{F}_\nu {}^\lambda  \,,
\ee
where $\tilde{\Gamma}^\lambda {}_{\rho\sigma} $ denotes the affine connection defined by the metric $\tilde{g}_{\mu\nu}$. Here we have chosen a gauge for the shifting symmetry~(\ref{homothety}) such that $\tilde{g}_{\lambda\nu} \dot{X}^\lambda \dot{X}^\nu = -1$.
Notice that we consider the binary STU black holes carrying a small amount of charges ($Q\ll M$). To be specific, we consider the STU black holes with a single charge (KK), two equal charges (EMDA), and four equal charges (KN). Substituting the solutions (\ref{kksol}), (\ref{emdasol}) and (\ref{knsol}) into above equation, and expressing the resulting equation in terms of the physical mass $M$ and charge $Q$, one can obtain the same standard radial equation of motion
\be
\mu \left(\frac{d^2 r}{dt^2} +\frac{M}{r^2}  \right) = \frac{q Q}{r^2} \,,
\ee
for different values of $\beta$ by imposing the Newtonian limit conditions. This can be seen as the equation of motion  of two interacting charged massive particles
\be
\frac{M_1 M_2}{M_1 +M_2} \frac{d^2 r}{d t^2} +\frac{M_1 M_2}{r^2} = \frac{Q_1 Q_2}{r^2} \,,
\ee
where $\mu = M_1 M_2/M $ and $ q = Q_1 Q_2/Q$ can be seen as the reduced mass and the charge of the test particle.  we have now identified the mass and charge assignments for the BKL recipe by imposing the scaling symmetry in supergravities.

\section{Final spin estimation} \label{sec4}

Now we apply the BKL recipe to estimate the final spin of merger of a binary rotating STU  black hole with different charge configurations and  different initial spin configurations, i.e.,  equal initial spins, unequal initial spins, and general initial spins. As the first step in exploring whether the revised method could improve the precision of estimation, we will study the final spin's difference between two cases corresponding to whether the scaling symmetry is taken into account or not.

\subsection{ISCO}

\begin{figure}[htbp]
\centerline{
 \includegraphics[width=0.33\linewidth]{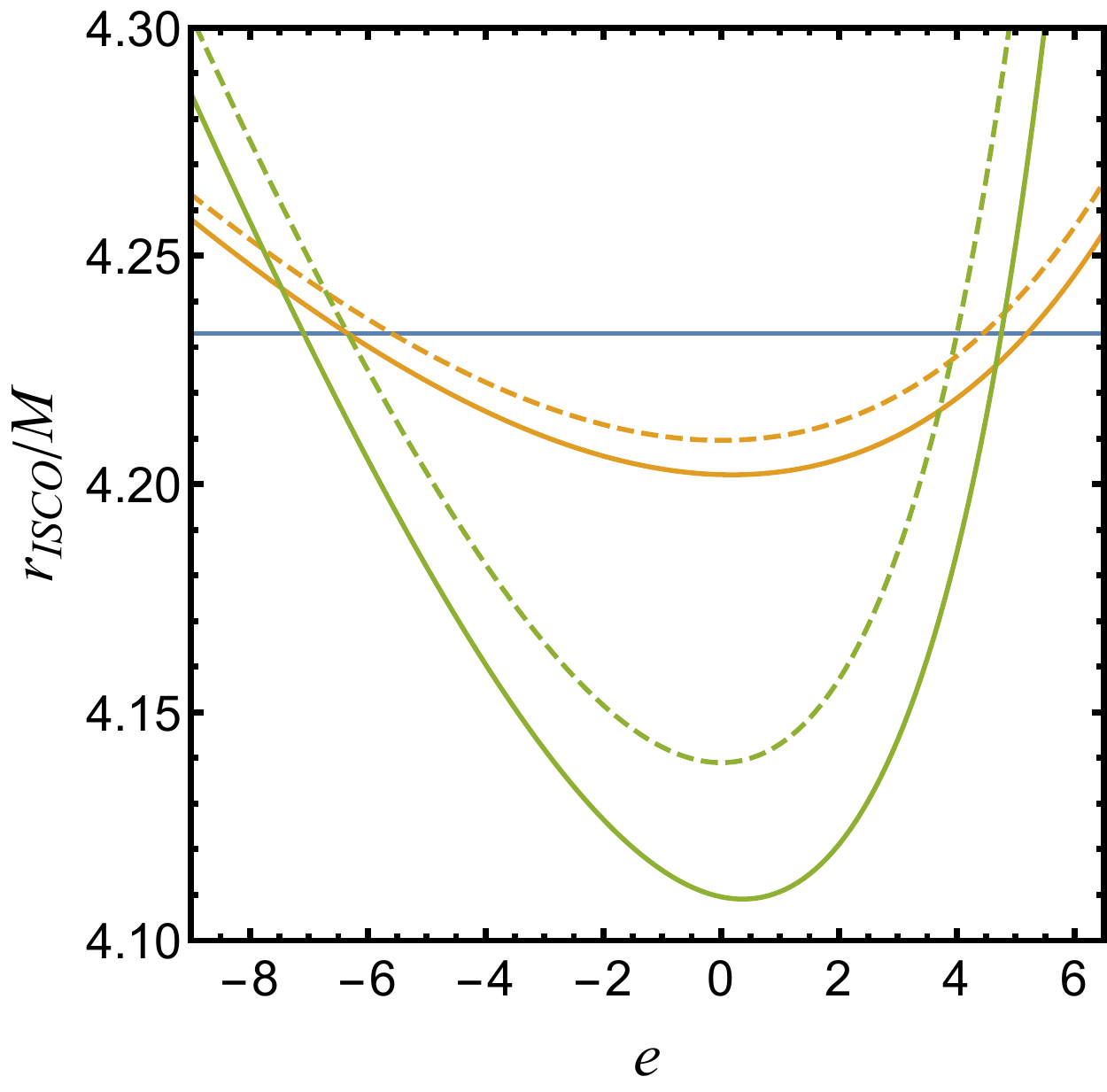} \   \includegraphics[width=0.33\linewidth]{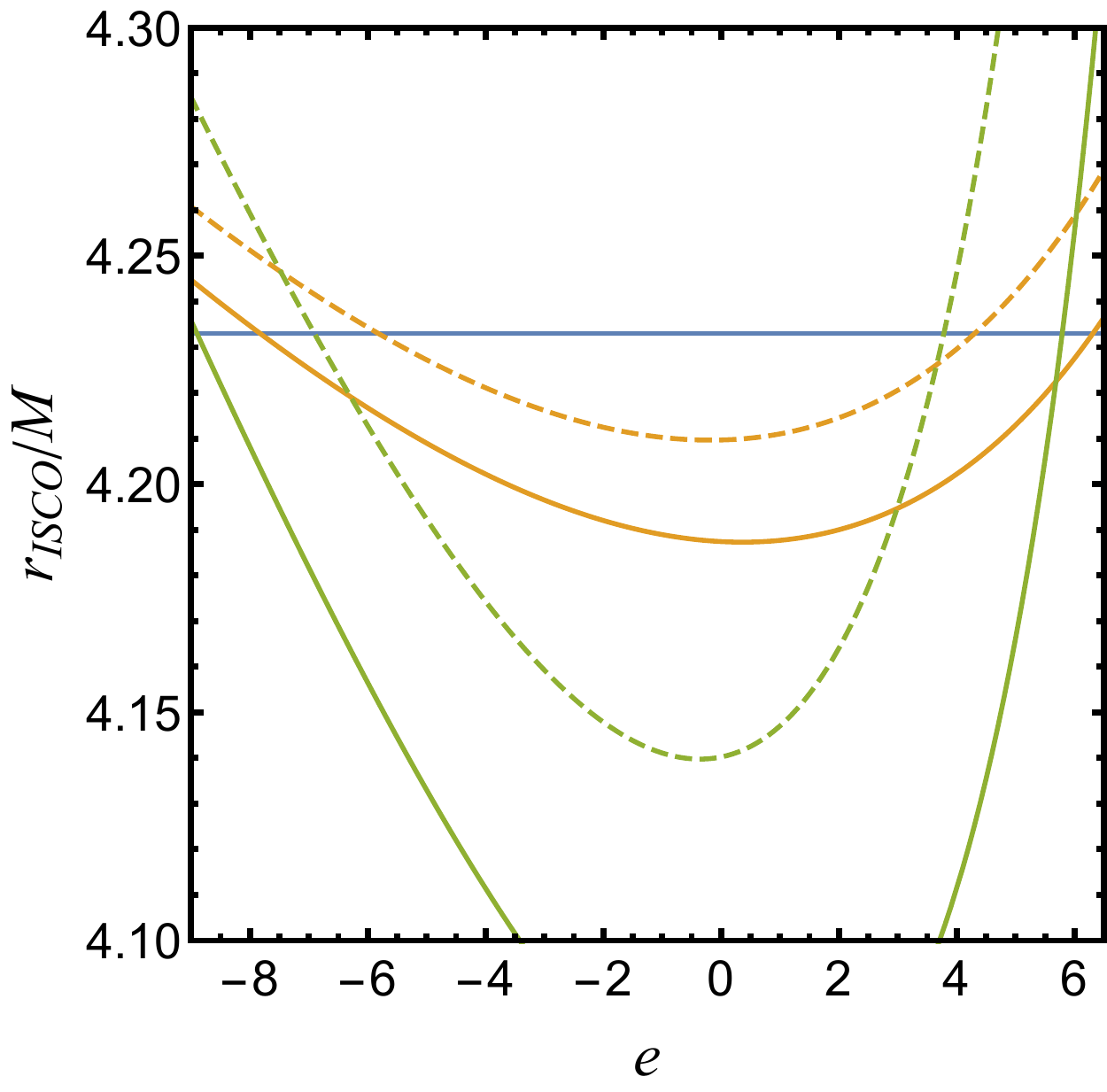} \
 \includegraphics[width=0.33\linewidth]{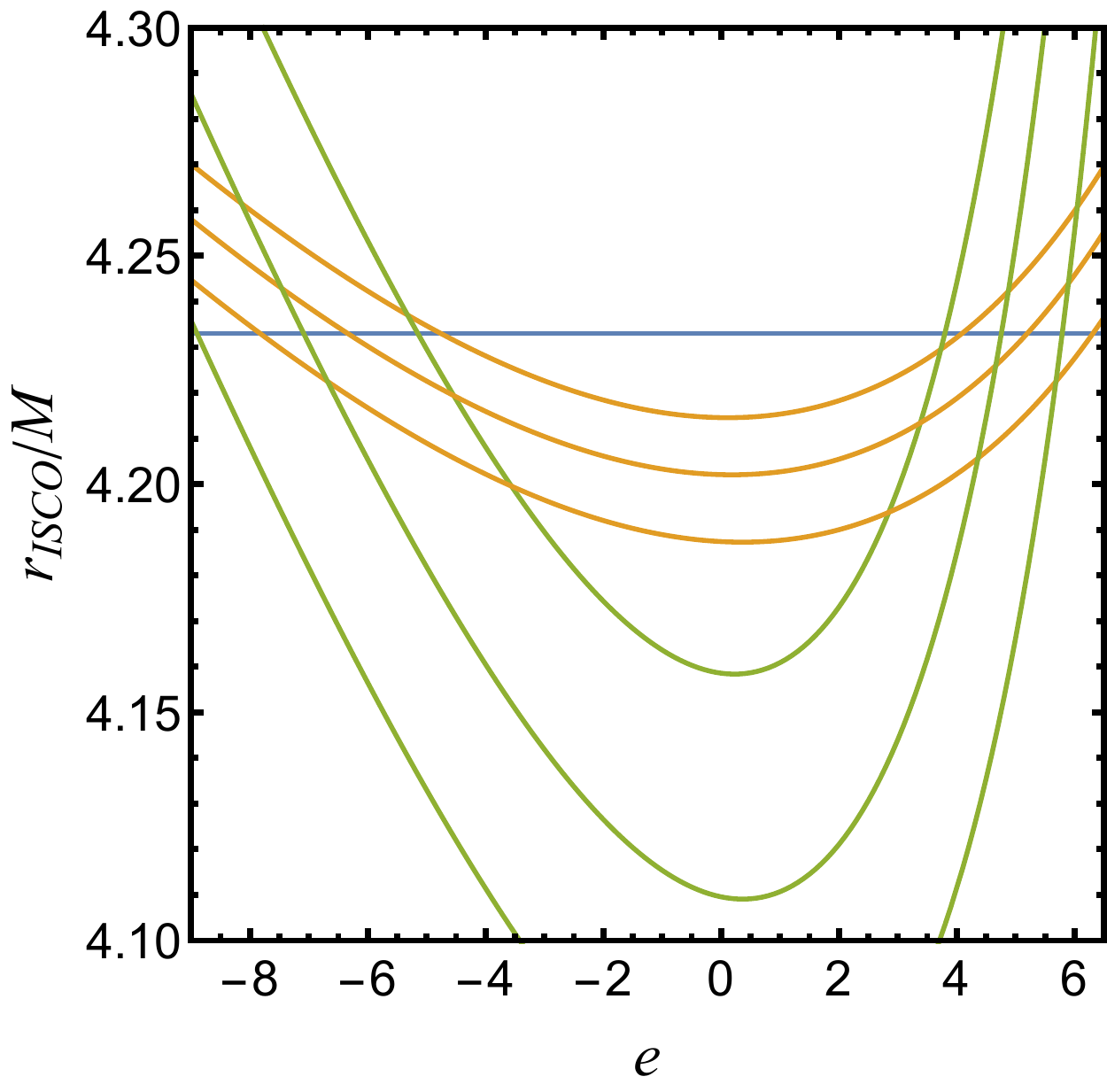}
 }
\caption{The radius $r_{\textup{ISCO}}/M$ against the charge to mass ratio $e$ in different cases. We set $A_f =0.5 M$ for simplicity. The blue, orange and green lines represent the $Q=0, Q= 0.05 M$, and $ Q = 0.1 M$ cases respectively. The  solid lines represent the cases in which the Lagrangian of test particle preserves the scaling symmetry, and the dashed lines represent the  cases in which the symmetry is not imposed. Left: EMDA; middle: KK; right: the lines in same color represent the KN, EMDA and KK cases respectively from top to bottom. }
\label{rvse0}
\end{figure}

\begin{figure}[htbp]
\centerline{
 \includegraphics[width=0.33\linewidth]{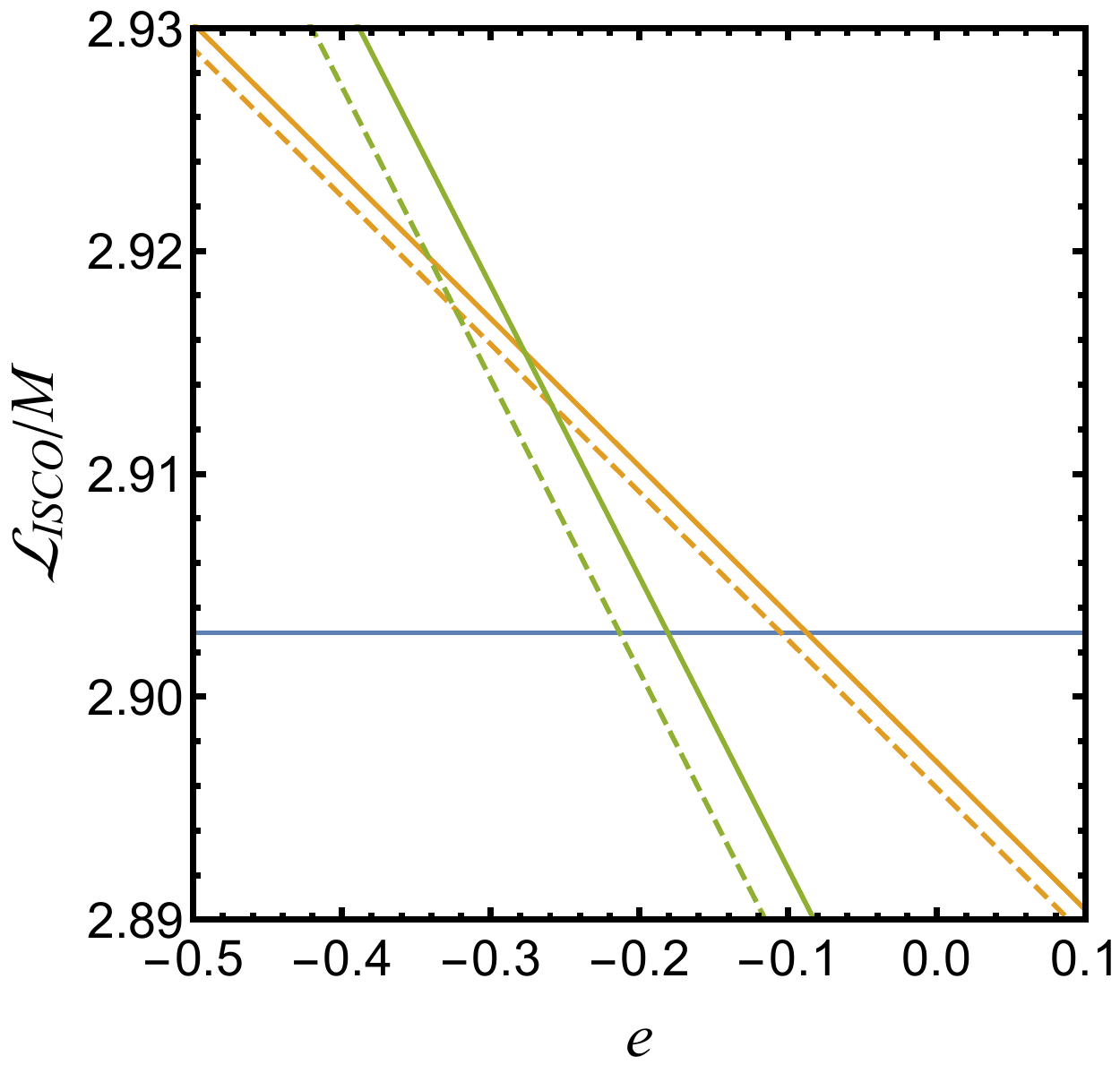} \   \includegraphics[width=0.33\linewidth]{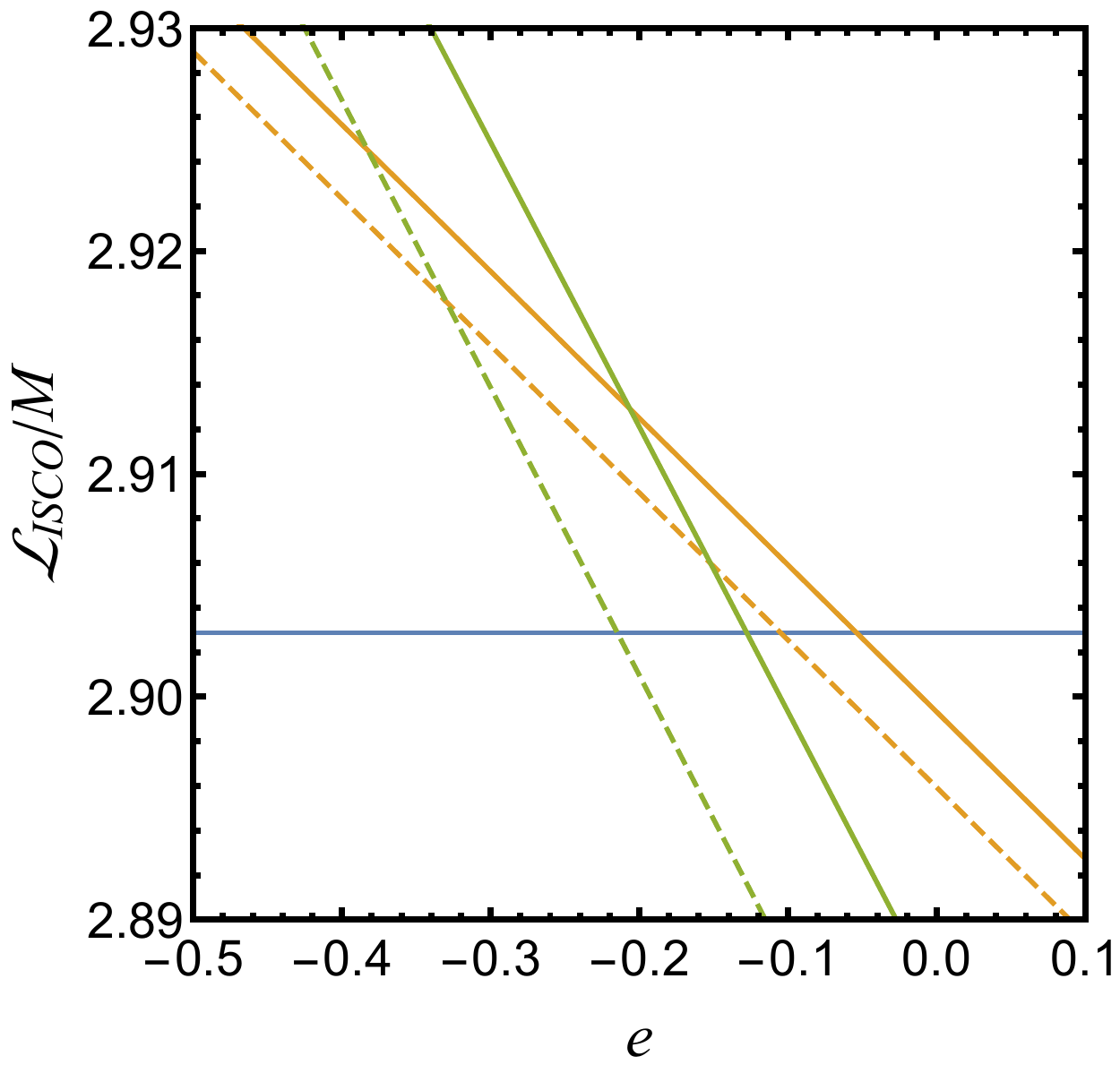} \
  \includegraphics[width=0.33\linewidth]{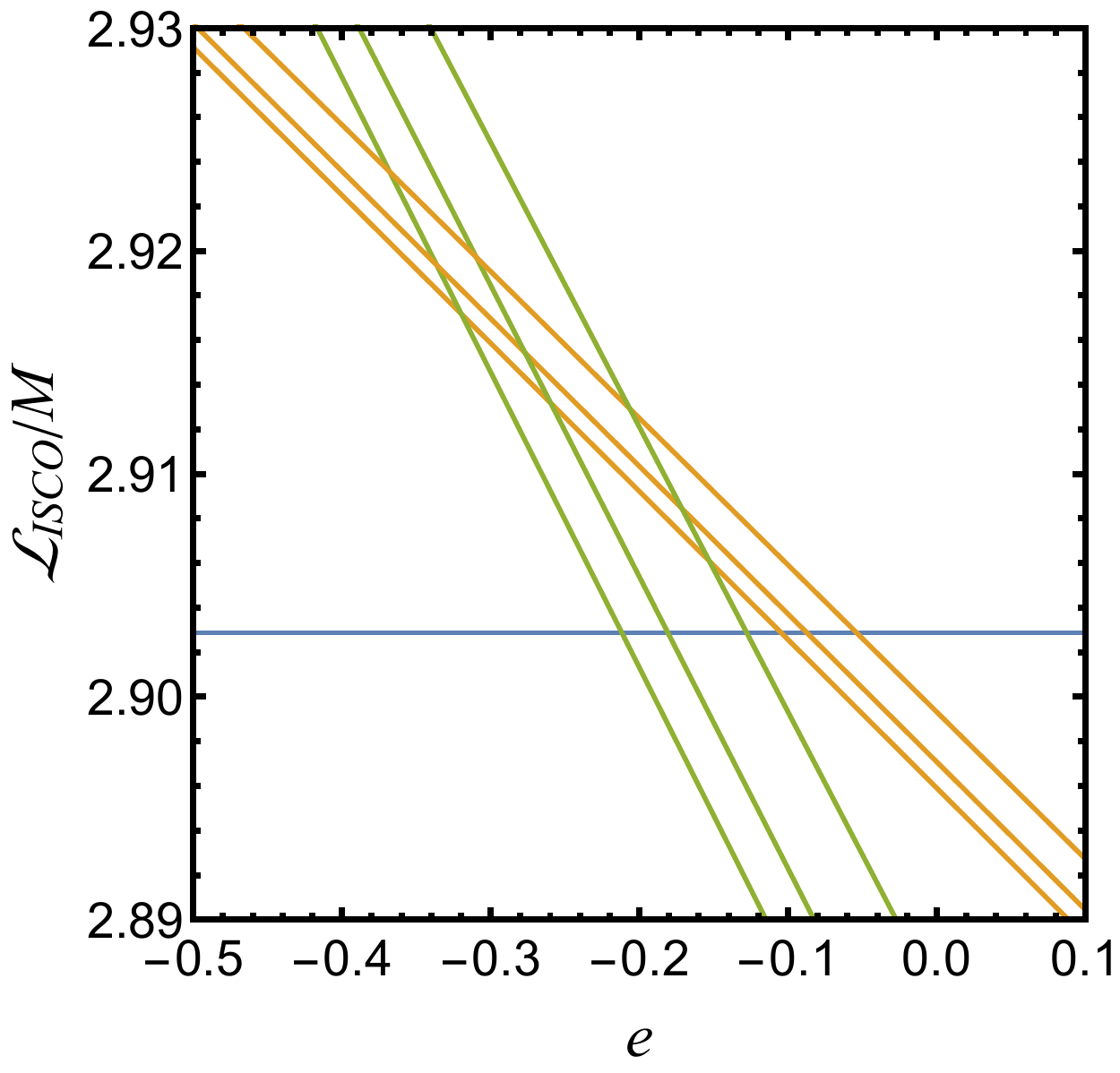}
 }
\caption{The angular momentum of the test particle with unit mass ${\cal L}_{\textup{ISCO}}/M$ against the charge to mass ratio $e$ in different cases. We set  $A_f =0.5 M$. The blue, orange and green lines represent the $Q=0, Q= 0.05 M$, and $ Q = 0.1 M$ cases respectively. The  solid lines represent the case that the Lagrangian of test particle preserves the scaling symmetry, and the dashed line represent the  case that the symmetry is not imposed. Left: EMDA; middle: KK; right: the lines in same color represent the KN, EMDA and KK cases respectively from left to right. }
\label{lvse0}
\end{figure}

As mentioned before, the particle motion will be modified if the scaling symmetry is taken into consideration. We will  study the ISCO which is related to the the test particle motion.
The conjugate momentum with respect to  $X^\lambda$ is given by
\be
P_\lambda = \frac{\partial L_2}{\partial \dot{X}^\lambda} = \frac{\mu \tilde{g}_{\lambda \nu} \dot{X}^\nu}{\sqrt{ -\tilde{g}_{\lambda\nu} \dot{X}^\lambda \dot{X}^\nu }} -\frac14 q A_\lambda = \mu \tilde{g}_{\lambda \nu} \dot{X}^\nu -\frac14 q A_\lambda \,,
\ee
where we have used the gauge
\be
\tilde{g}_{\mu\nu} \dot{X}^\mu \dot{X}^\nu = -1 = \tilde{g}_{tt} \dot{t}^2 +\tilde{g}_{rr} \dot{r}^2 +\tilde{g}_{\theta\theta} \dot{\theta}^2 +\tilde{g}_{\phi\phi} \dot{\phi}^2 +2 \tilde{g}_{t\phi} \dot{t} \dot{\phi}\,, \label{gauge}
\ee
Now we consider the motion of a charged massive particle in the equatorial plane of the KK rotating
black hole, determined by $\theta = \pi/2$ and $\dot{\theta} =0$. The enery $ \cal E$ and angular momentum $\cal L$ of the test particle with unit mass are given by
\bea
\cal E &=& -\frac{P_t}{\mu}  = - \tilde{g}_{tt} \dot{t} -\tilde{g}_{t\phi} \dot{\phi} +\frac14 e A_t \,,\\
\cal L &=& \frac{P_\phi}{\mu}  = \tilde{g}_{\phi\phi} \dot{\phi} +\tilde{g}_{t\phi} \dot{t} -\frac14 e A_\phi \,,
\eea
where $e = q/\mu$ represents the charge to mass ratio of the test particle. We can obtain
\bea
\dot{t} &=& \frac{(4 {\cal L}+e A_\phi) \tilde{g}_{t\phi} +(4 {\cal E} - e A_t ) \tilde{g}_{\phi\phi} }{4 \tilde{\Delta}_r} \,, \label{tdot} \\
\dot{\phi} &=& -\frac{(4 {\cal L}+e A_\phi) \tilde{g}_{tt} +(4 {\cal E} - e A_t ) \tilde{g}_{t\phi} }{4 \tilde{\Delta}_r} \,, \label{phidot}
\eea
where $ \tilde{\Delta}_r = \tilde{g}_{t \phi}^2 - \tilde{g}_{tt} \tilde{g}_{\phi\phi}$.
Substituting Eqs.~(\ref{tdot}) and (\ref{phidot}) into (\ref{gauge}), we can define the effective potential
\be
V_{\textup{eff}} \equiv  \dot{r}^2 = \frac{(4 {\cal L} +e A_\phi)^2 \tilde{g}_{tt} + (4 {\cal E} -e A_t)^2 \tilde{g}_{\phi\phi} +2(4 {\cal L} +e A_\phi)({4 \cal E} -e A_t) \tilde{g}_{t\phi} -16 { \tilde{\Delta}_r}}{16 \tilde{\Delta}_r \tilde{g}_{rr}} \,.
\ee
The ISCO can be found by imposing the following conditions
\be
V_{\textup{eff}} =0 \,, \quad V^{\prime}_{\textup{eff}} = 0 \,, \quad V^{\prime\prime}_{\textup{eff}} =0 \,,
\ee
where the prime denotes the derivative with respect to $r$. We  plot the $r_{\textup{ISCO}}$ against $e$ in the EMDA and KK  cases in Fig.~\ref{rvse0}.
First, as the charge to mass ratio $e$ varies from negative to positive, the ISCO radius gradually reduces to a minimal value, and then begins to increase if the black hole carries nonzero charge. Second,  we find that the ISCO radius will be smaller when we  require the test particle to preserve the scaling symmetry in both KK and EMDA cases. As the black hole carries more charges, the difference  becomes larger of the ISCO radius between two cases corresponding to whether the symmetry is taken into account or not. From the right plot of Fig.~\ref{rvse0}, we observe that for the final STU black hole carrying a certain number of charges, the ISCO radius of the orbiting test particle gets smaller in the order of an STU black hole with four equal charges (KN), two equal charges (EMDA) and a single nonzero charge (KK).  One can also understand the variation of the ISCO radius as due to the variation of the coupling constant $\beta$ from zero to $\sqrt{3}$.
We also plot the orbital angular momentum ${\cal L}_{\textup{ISCO}}/M$ against $e$ in EMDA  and KK  cases in Fig.~\ref{lvse0}.
First, as the charge to mass ratio varies from negative to positive, the angular momentum gradually decreases if the black hole carries nonzero charge. This can be understood physically as a result of the fact that an attractive electric force helps to increase the angular momentum while a repulsive one does the opposite. Second, we find that the angular momentum will be larger if the test particle is required to preserve the scaling symmetry in both KK  and EMDA  cases. As the final black hole carries more charges, the difference of the angular momentum between two cases corresponding to whether the symmetry is taken into account or not  becomes  larger. From the right plot of Fig.~\ref{lvse0}, we can see that for the final black hole that carries a certain number of charges, the angular momentum of the orbiting test particle  becomes smaller in the order of STU black hole with four equal charges, two equal charges and a single nonzero charge.  One can also attribute this decrease of the orbital angular momentum  to the increase of the coupling constant $\beta$ from zero to $\sqrt{3}$.

\subsection{Equal initial spins }

\begin{figure}[htbp]
\centerline{
 \includegraphics[width=0.45\linewidth]{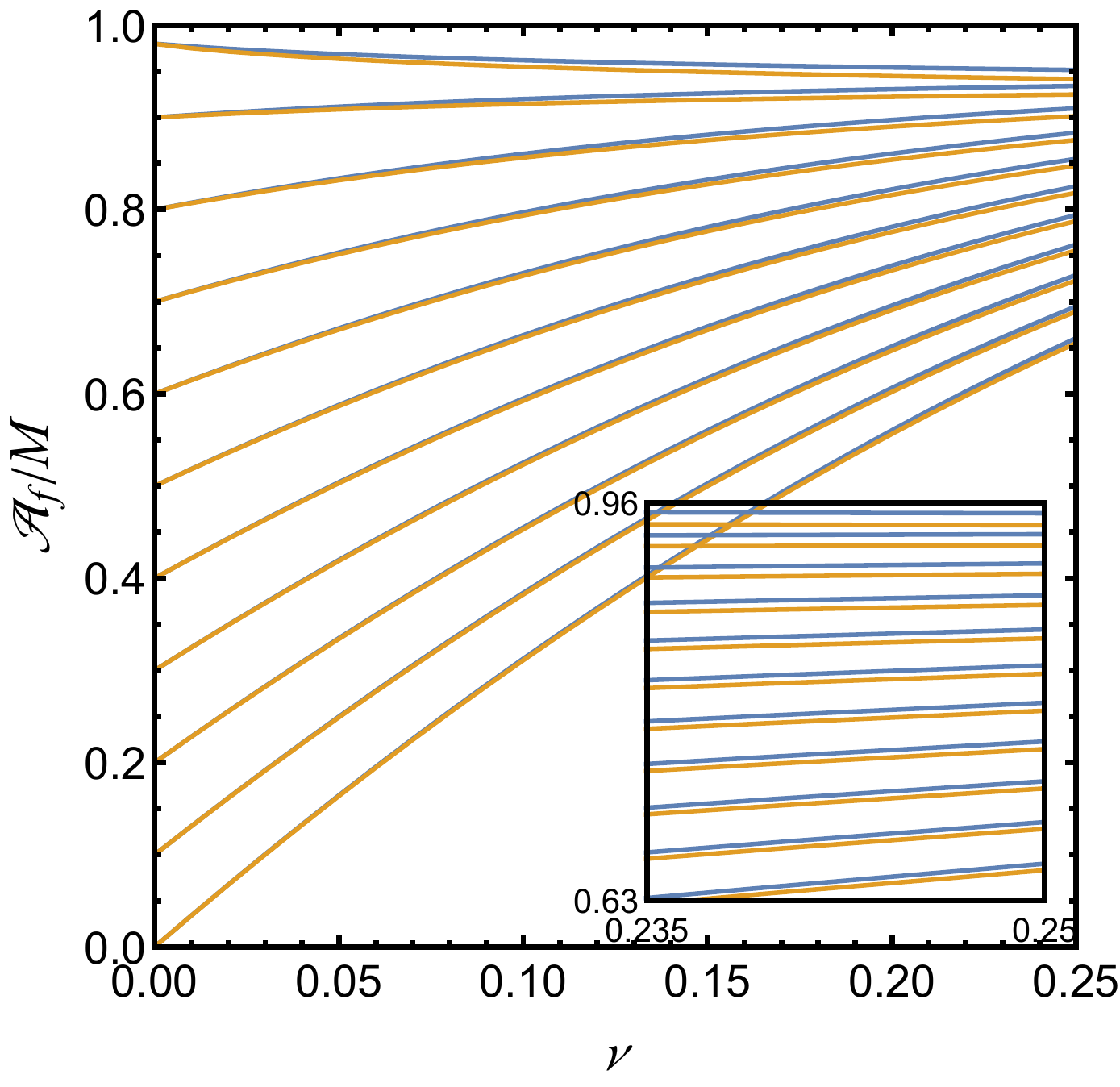}\ \includegraphics[width=0.45\linewidth]{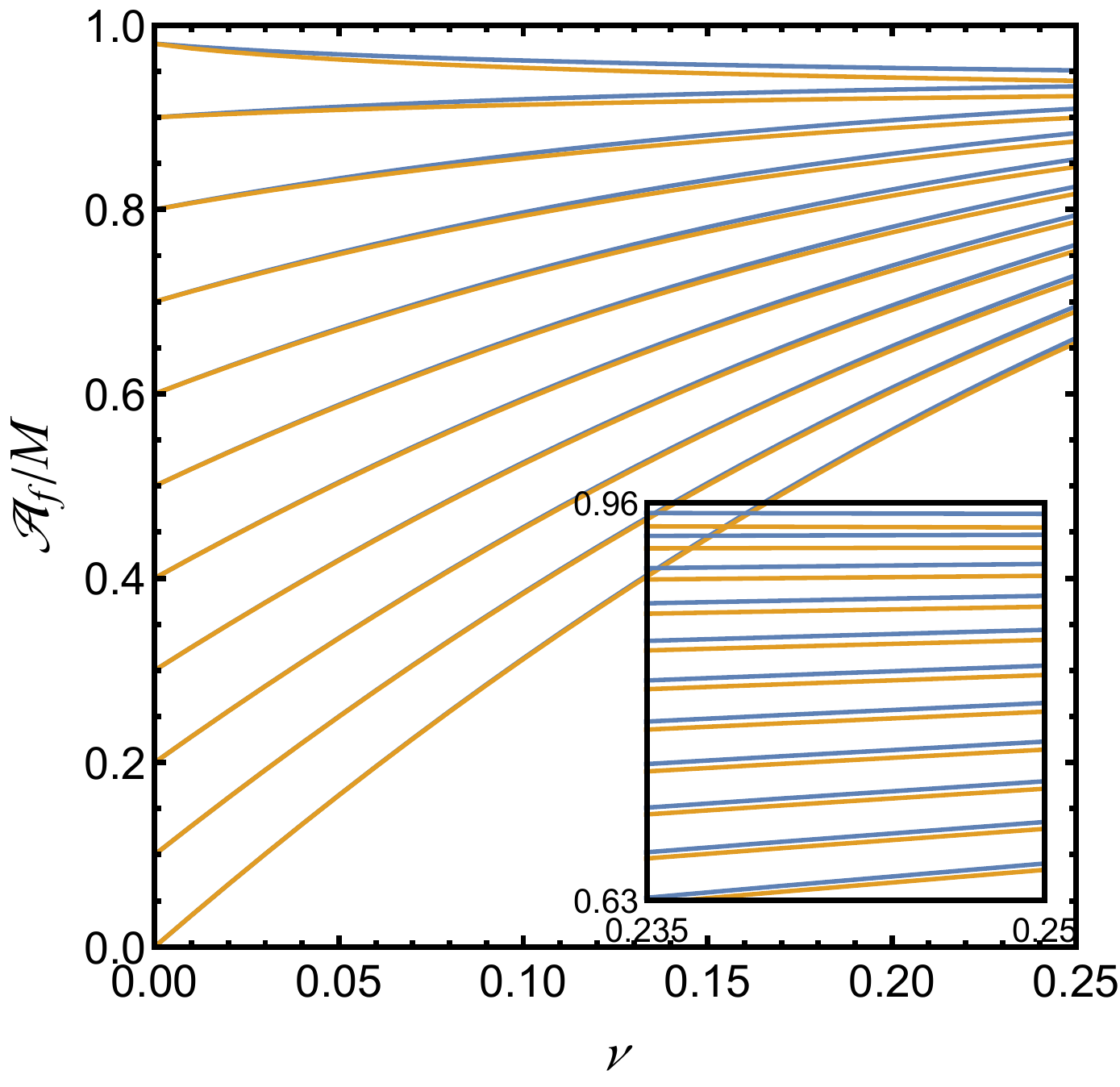}
}
\centerline{
 \includegraphics[width=0.45\linewidth]{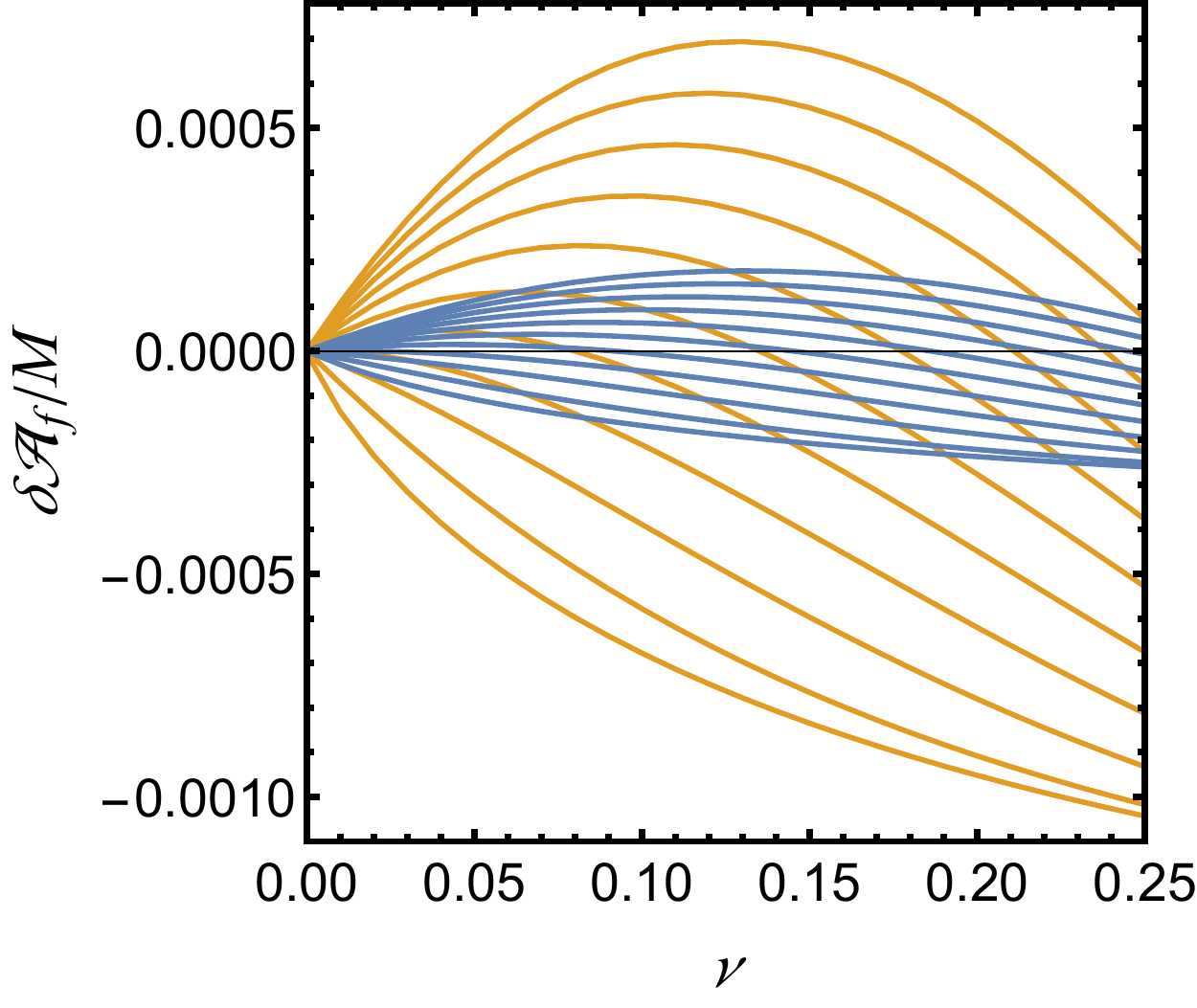} \
\includegraphics[width=0.45\linewidth]{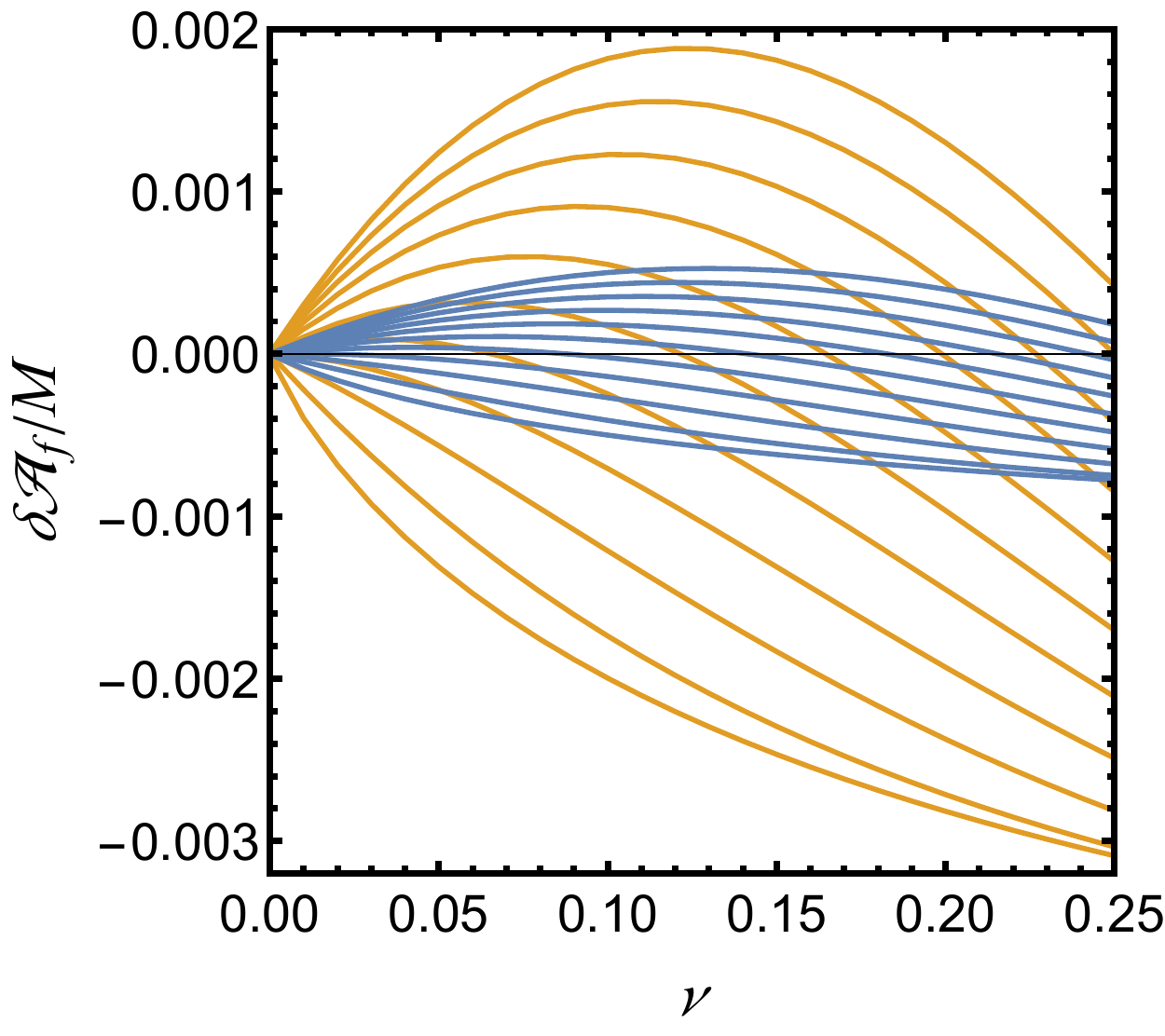}
}
\caption{ The blue  and orange lines represent  $Q= 0.05 M$ and $ Q = 0.1 M$ cases respectively. We set $e =0.1$. Top: The final spin of the remnant black hole with unit mass against $\nu$ in the EMDA (left) and KK (right) cases.
The initial spin $\chi$ of the lines in same color is 0, 0.1, 0.2, 0.3, 0.4, 0.5, 0.6, 0.7, 0.8, 0.9, 0.98 respectively from bottom to top.
Bottom: the final spin's differences $\delta {\cal A}_f$ of  remnant black holes between the two cases corresponding to the symmetry is taken into account or not in EMDA (left) and KK (right) cases. The initial spin $\chi$ of the lines in same color is 0, 0.1, 0.2, 0.3, 0.4, 0.5, 0.6, 0.7, 0.8, 0.9, 0.98 respectively from top to bottom. }
\label{Afnu1}
\end{figure}

Now we apply the BKL recipe to estimate the final spin. In this subsection, we assume that the initial spins of the BBH are equal, i.e. $\chi_1 = \chi_2 =\chi$. According to Eq.~(\ref{finalspin}), the final spin can be rewritten as
\be
{\cal A}_f  = {\cal L} \nu +M (1 -2 \nu) \chi \,.
\ee
Because we consider the merger of a BBH with weak charge, it is natural to consider the test particle carrying weak charge too. We set  $e =0.1$ for simplicity. (actually the test particle carrying different charges has similar behaviors, and so we do not consider the effect of the value of $e$ on the final spin.) Given $Q, \chi$ and $\nu$, we can solve the final spin.
We plot  ${\cal A}_f$ against $\nu$ for different initial spins $\chi$ from zero to 0.98 ($\chi=0$ and $\chi = 0.98$ represent a non-spinning black hole and  a near extreme black hole, i.e. a rapidly-spinning black hole, respectively) and different charges $Q = 0.05 M$ and $0.1 M$ in Fig.~\ref{Afnu1} respectively.
Notice that $\nu \sim0$ and $0.25$ represent the extreme initial mass ratio and equal initial masses respectively.

From the top plots in Fig.~\ref{Afnu1}, we find that some features of the final spin estimated by the BKL recipe while the symmetry is imposed in the binary charged black hole merger case are similar to that in the neutral case~\cite{Buonanno:2007sv} and charged case in which the symmetry is not taken into account~\cite{Jai-akson:2017ldo}.
First, the largest final spin for the remnant black hole is achieved for  a  binary extreme black hole merger with extreme initial mass ratio, which can be viewed as a charged particle falling into a charged rapidly-spinning black hole, regardless of the amount of charge carried by black hole.
Besides, regardless of the amount of charges carried by the binary system, the final spin for the merger of binary non-spinning charged black holes with extremely high mass ratios  is zero.
Third, there is a critical value for the initial spins,  below which, as the mass ratio approaches  that of the equal-mass case ($\nu =0.25$), the final spin increases. And above the critical value, the final spin decreases as  $\nu$ increases.
Fourth,  the final spin will be smaller as the BBH system carries more charges.

From the bottom plots in Fig.~\ref{Afnu1}, we find some subtle differences between two cases corresponding to whether the symmetry is taken into account or not.
First, there is no difference between the two cases if the initial mass ratio is extreme.   Second, while the initial black holes are  non-spinning, the final spin estimated by the BKL recipe in which the symmetry is imposed is always larger than that in which the symmetry is not taken into account. The difference  increases firstly and then decreases as the initial mass ratio approaches  unity. Third, there is  a threshold for the initial spins, above which
, the final spin estimated by the BKL recipe in which the symmetry is imposed is always smaller than that in which the symmetry is not taken into account.
All these features exist in different charge configurations of STU supergravity (both the KK and EMDA cases). It is worth comparing the final spin given by the BKL recipe with numeric simulations~\cite{Hirschmann:2017psw}, and exploring if the BKL recipe could provide more accurate prediction of the final spin by requiring that  the Lagrangian of the test particle preserves the scaling symmetry in supergravity.

\begin{figure}[htbp]
\centerline{
 \includegraphics[width=0.45\linewidth]{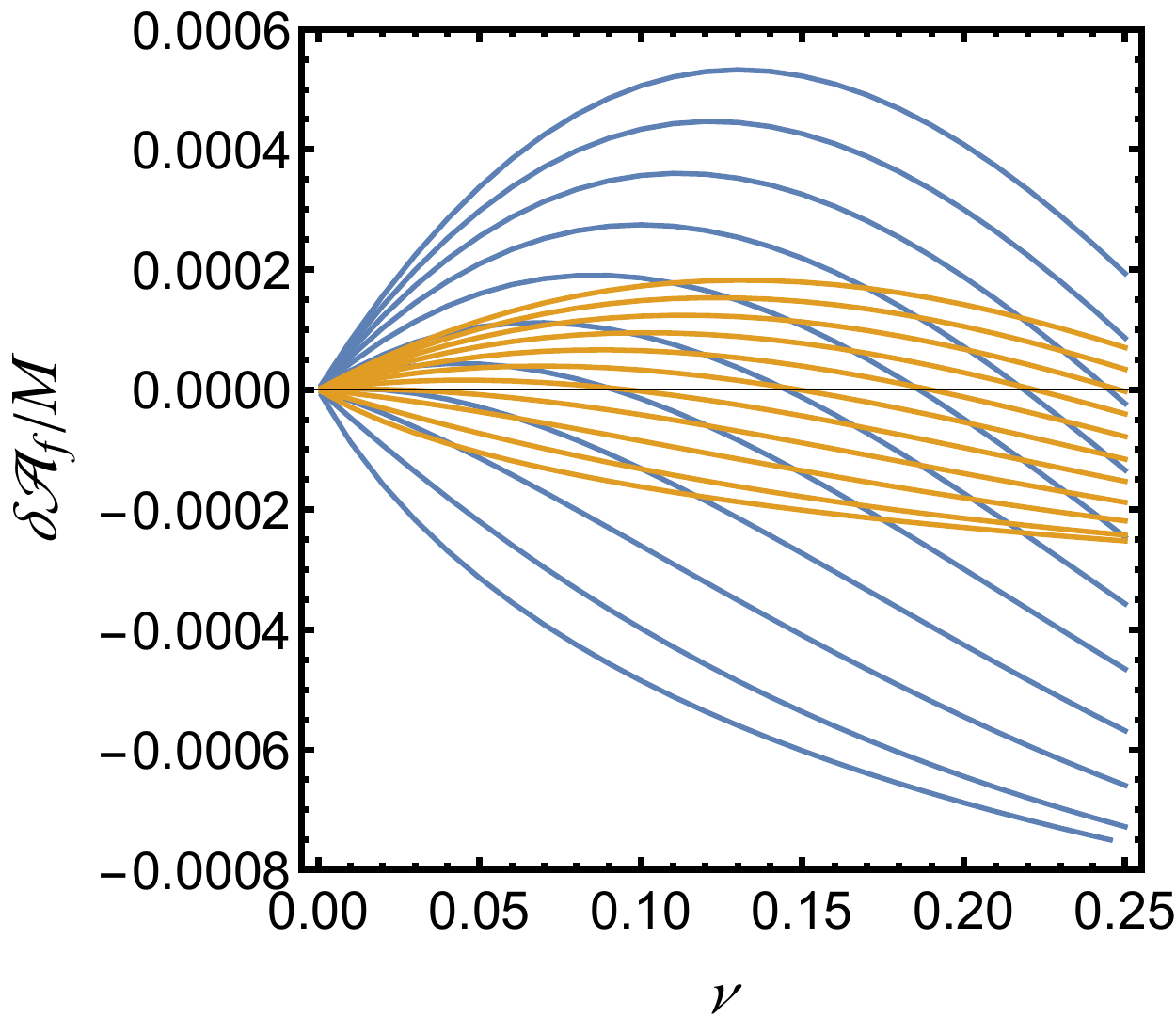} \
\includegraphics[width=0.45\linewidth]{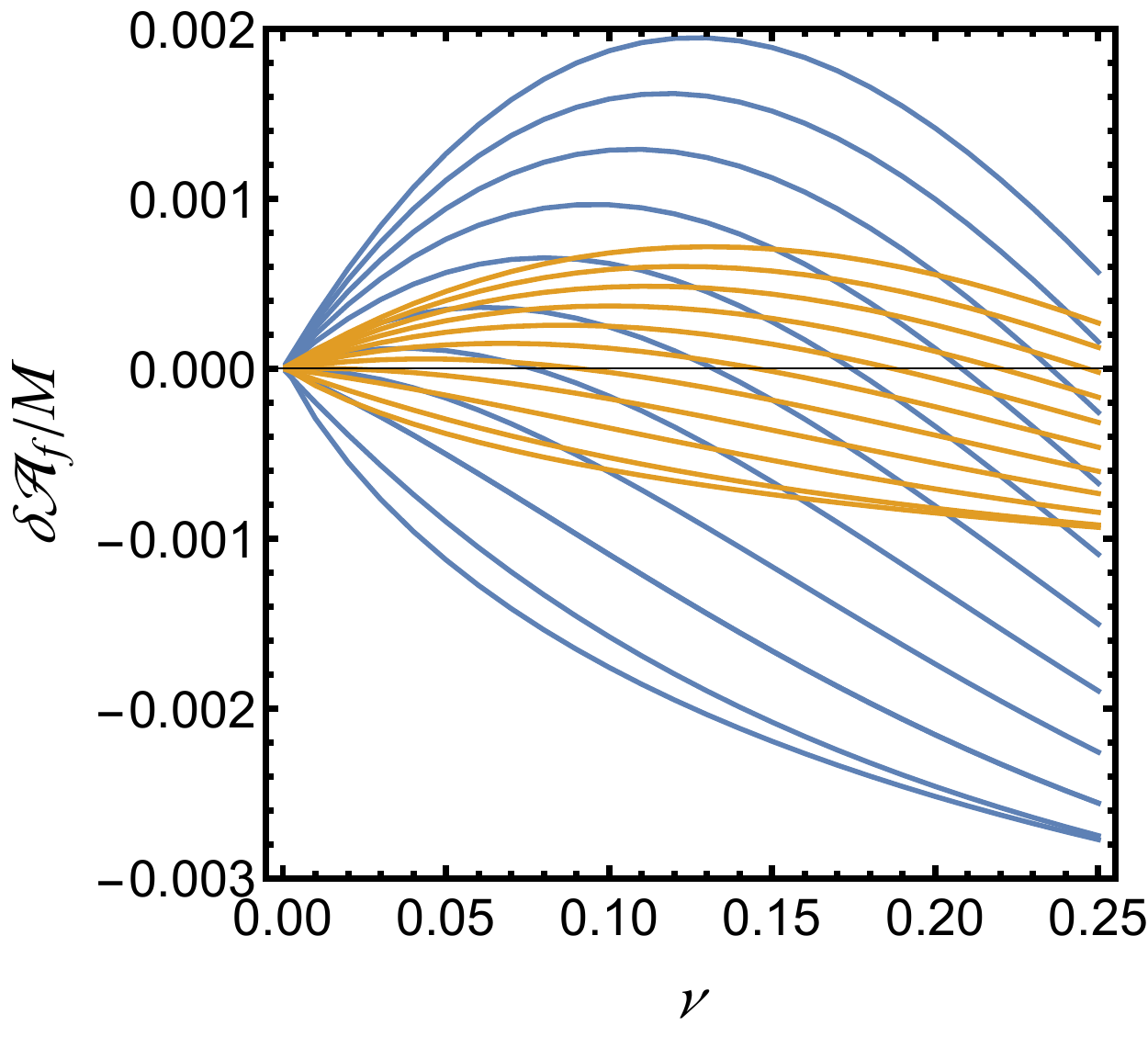}
}
\caption{Left: $Q= 0.05 M$; right: $ Q = 0.1 M$. The blue line represents the final spin's difference $\delta {\cal A}_f$ of the remnant KK black hole (STU black hole with single charge) and KN black hole (STU black hole with four equal charges). The orange line represents the final spin's difference $\delta {\cal A}_f$ of the remnant EMDA black hole (STU black hole with two equal charges) and KN black hole (STU black hole with four equal charges).  We set $e =0.1$.  The initial spins $\chi$ of the lines in same color are 0, 0.1, 0.2, 0.3, 0.4, 0.5, 0.6, 0.7, 0.8, 0.9 respectively from top to down.}
\label{Afnu1x}
\end{figure}

We also study the final spin's difference $\delta {\cal A}_f$ between different charge configurations of STU supergravity with a certain small number of charges. To be specific, we plot the final spin's difference between  the KK and KN black holes, and  between the EMDA and KN black holes in Fig.~\ref{Afnu1x}.
We find that the final spin's  differences between the cases of different charge configurations  have  similar features as that between two cases corresponding to whether the scaling symmetry is taken into account or not. As mentioned in previous subsection, the difference can be also explained by the coupling constant in STU supergravity. First, there is no difference between the cases of different charge configurations (different coupling constant) if the initial mass ratio is extreme. Second, while the initial spins are zero, the final spin in the case of the single charge configuration  (the coupling constant is $\sqrt{3}$) is always larger than that in the cases of other charge configurations (the coupling constant is smaller than $\sqrt{3}$). The difference  increases firstly and then decreases as the equal mass limit is approached.  Third, there is also a critical value for the initial spins, above which, the final spin in the case of the single charge configuration (the coupling constant is $\sqrt{3}$) is always smaller than that in the cases of other charge configurations (the coupling constant is smaller than $\sqrt{3}$). The difference decreases gradually as the initial mass ratio approaches unity.   All these results may provide a potential way to test different supergravites near strong gravitational regimes.  Let us note  that  the final spin's difference between the two cases corresponding to whether the symmetry is taken into account or not is small because we only study the merger of a BBH with weak charges according to previous astronomical observations. If we consider the merger of a BBH with a  greater amount of charges (we plot this case in the appendix~\ref{app2}), the difference would become large similar to the result in~\cite{Jai-akson:2017ldo}. We hope that the final spin's difference could be detected by future precise detectors, and could be used to test supergravities.

\subsection{Unequal  initial spins }

\begin{figure}[htbp]
\centerline{
 \includegraphics[width=0.33\linewidth]{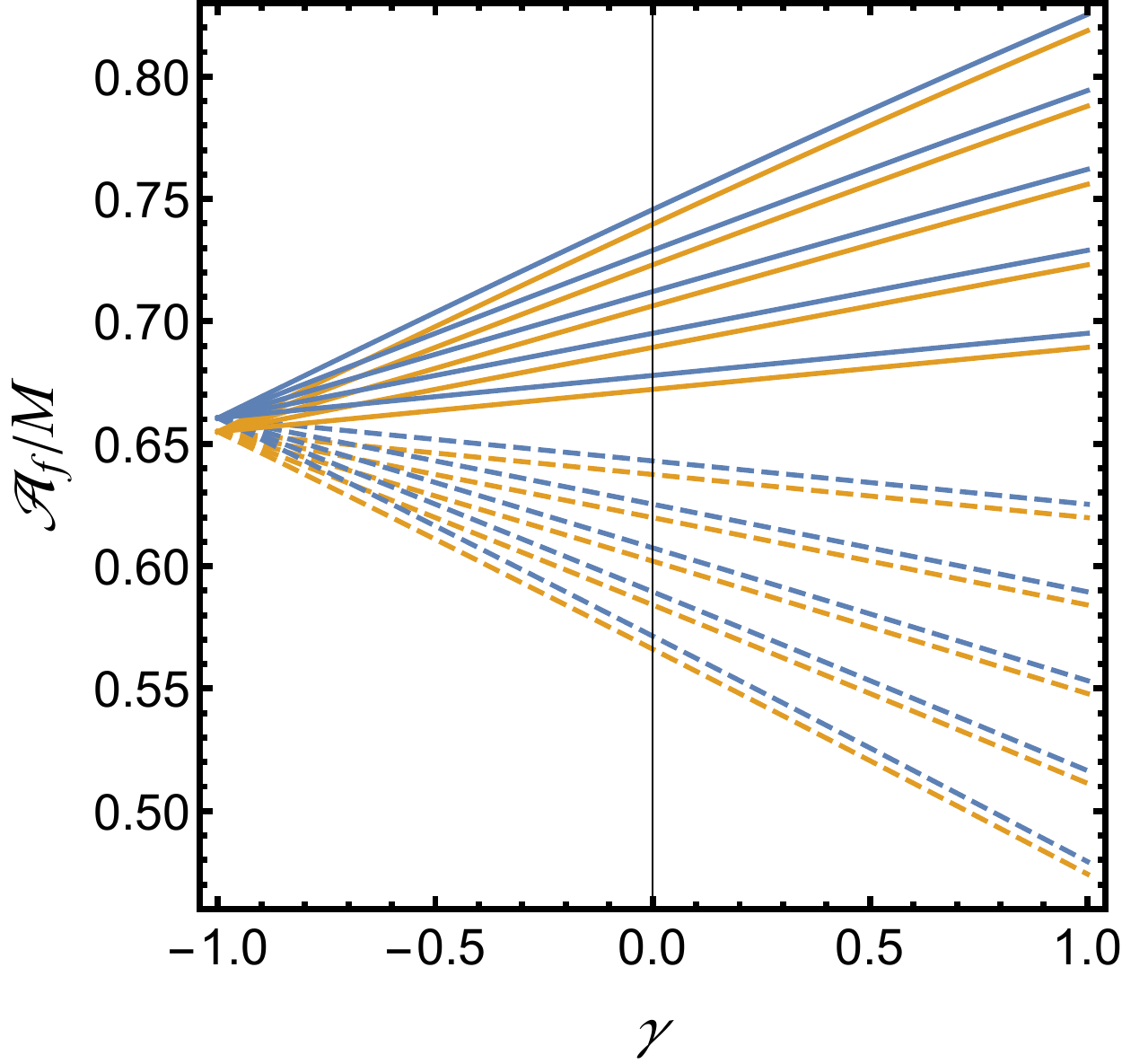} \
 \includegraphics[width=0.33\linewidth]{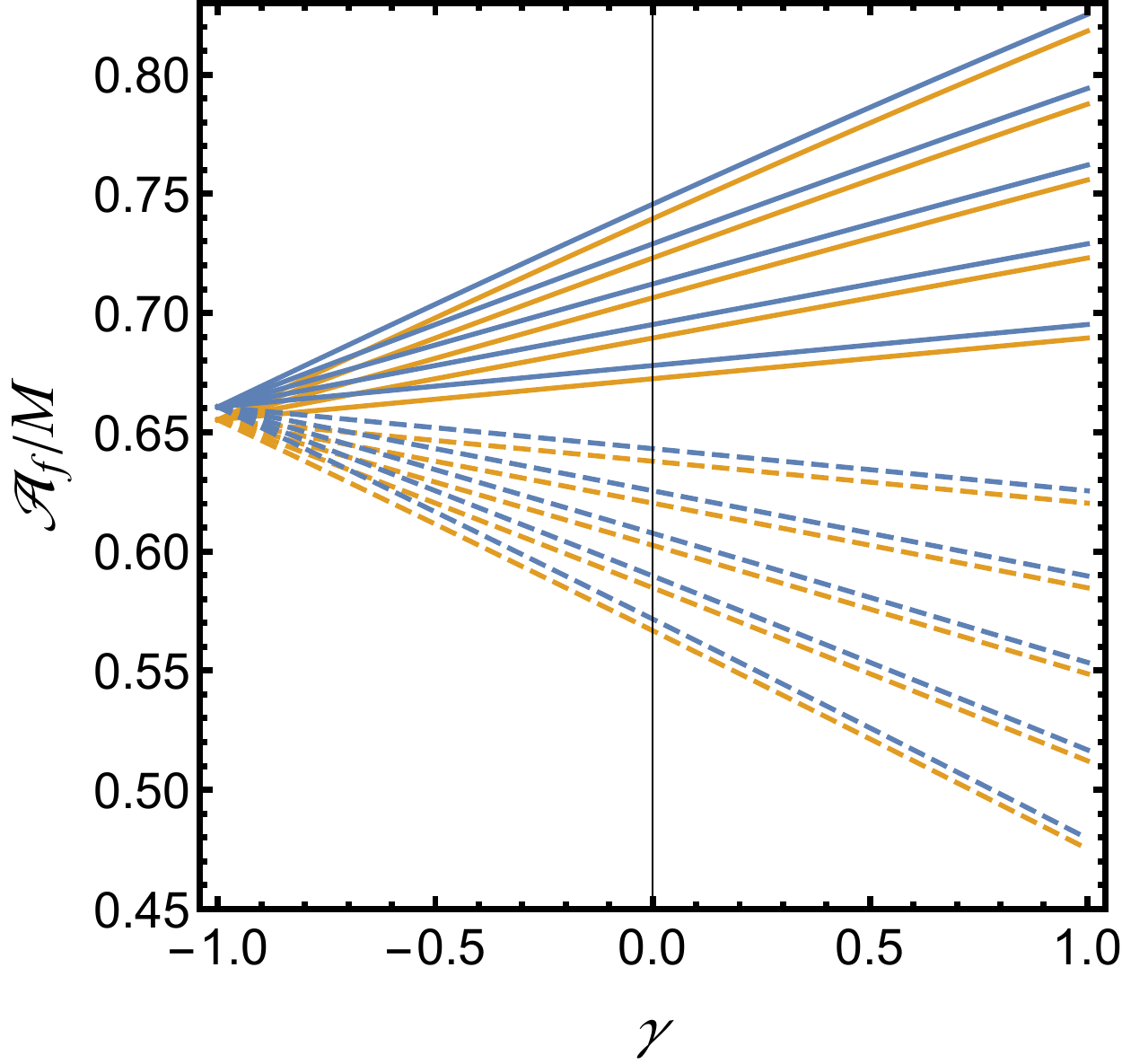} \
\includegraphics[width=0.33\linewidth]{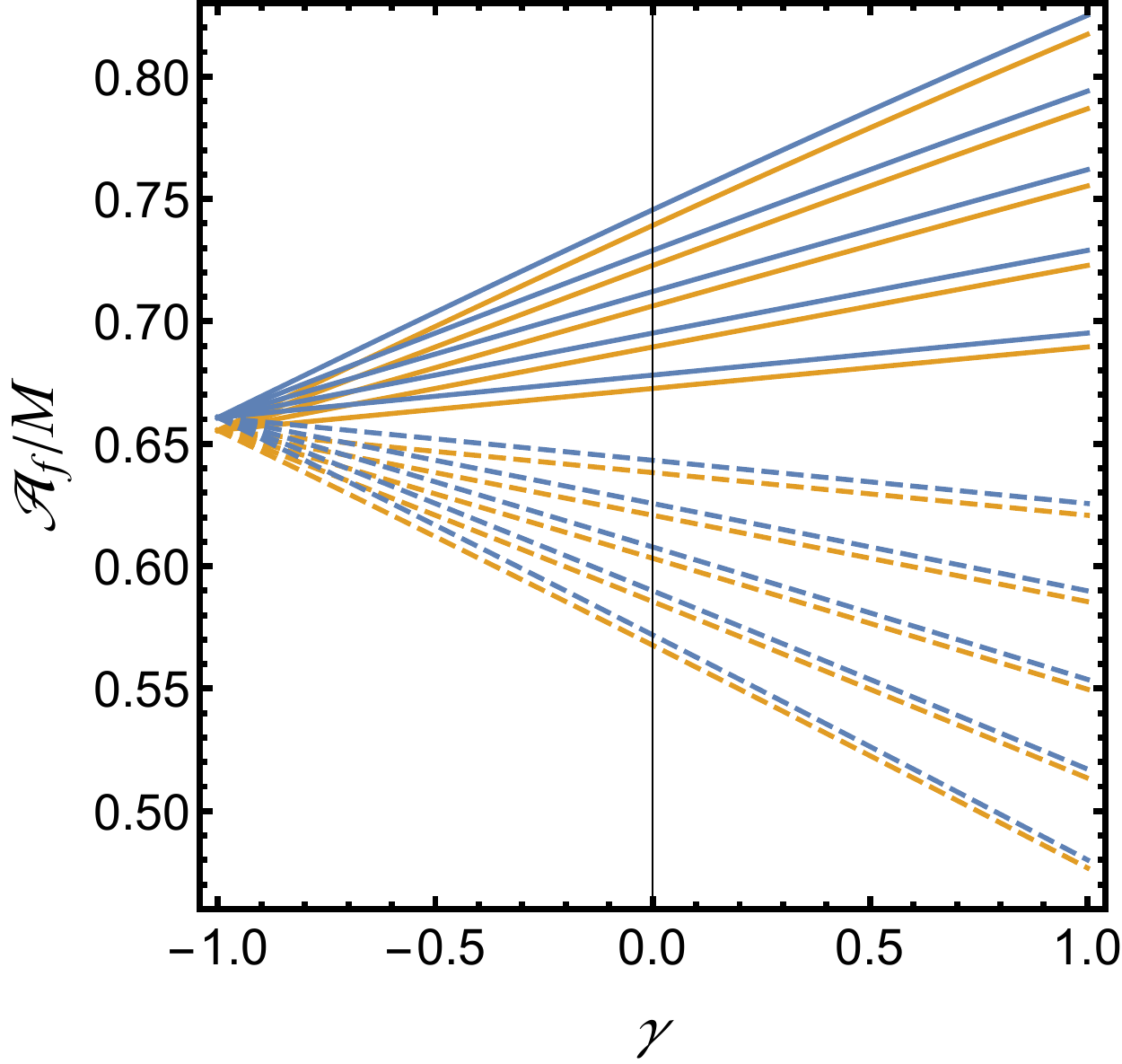}
}
\caption{Left: KN; middle: EMDA; right: KK. We set  $e =0.1$ for simplicity. The blue lines and orange lines represent $Q= 0.05 M$ and $ Q = 0.1 M$ cases respectively. The solid lines and dashed lines represent positive and negative initial spin $\chi$  respectively.  The value of $\chi$ of the lines in same color is 0.5, 0.4, 0.3, 0.2, 0.1, -0.1, -0.2, -0.3, -0.4, -0.5 respectively from top to bottom.}
\label{Afnu2}
\end{figure}

\begin{figure}[htbp]
\centerline{
 \includegraphics[width=0.33\linewidth]{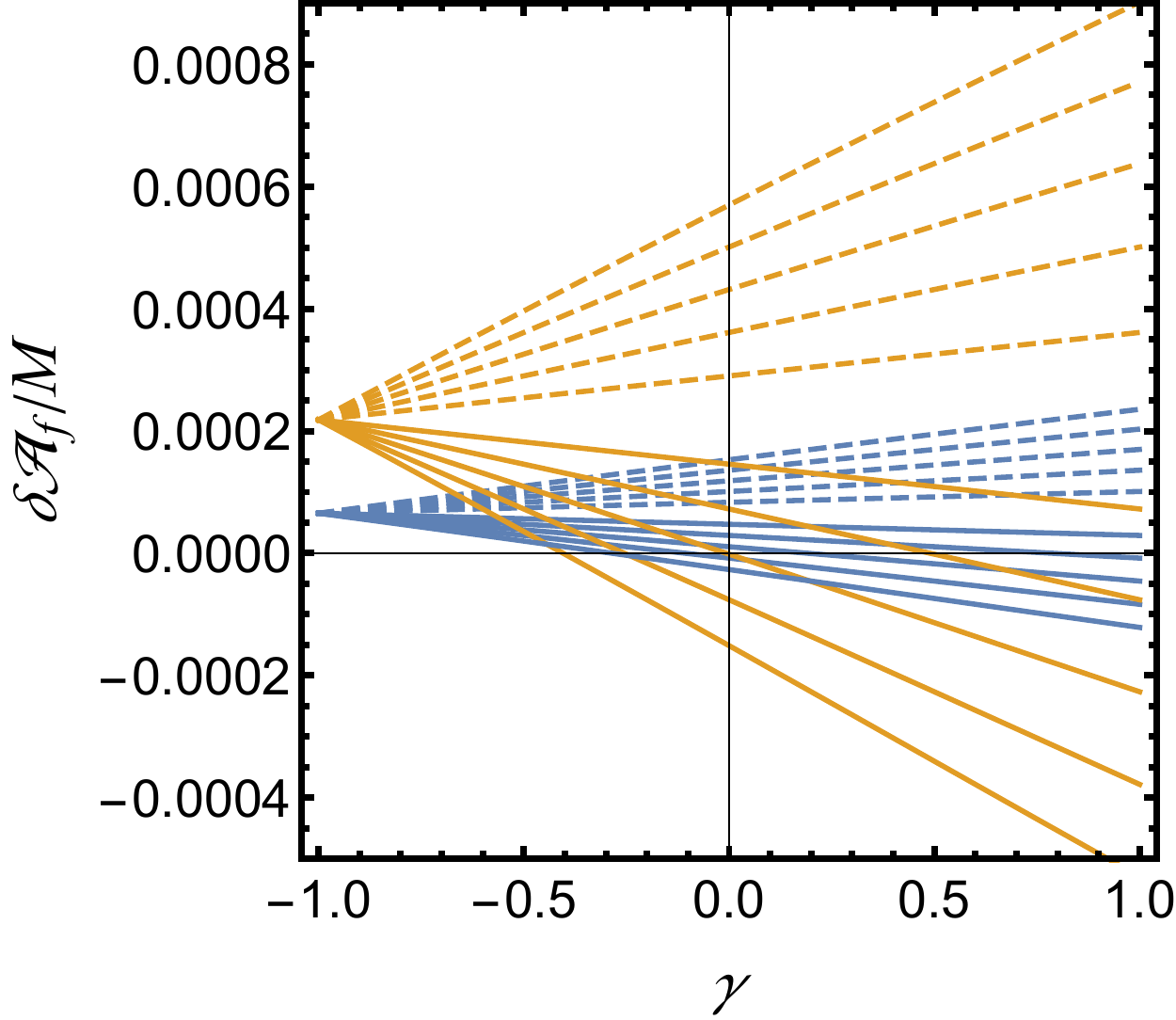} \
 \includegraphics[width=0.33\linewidth]{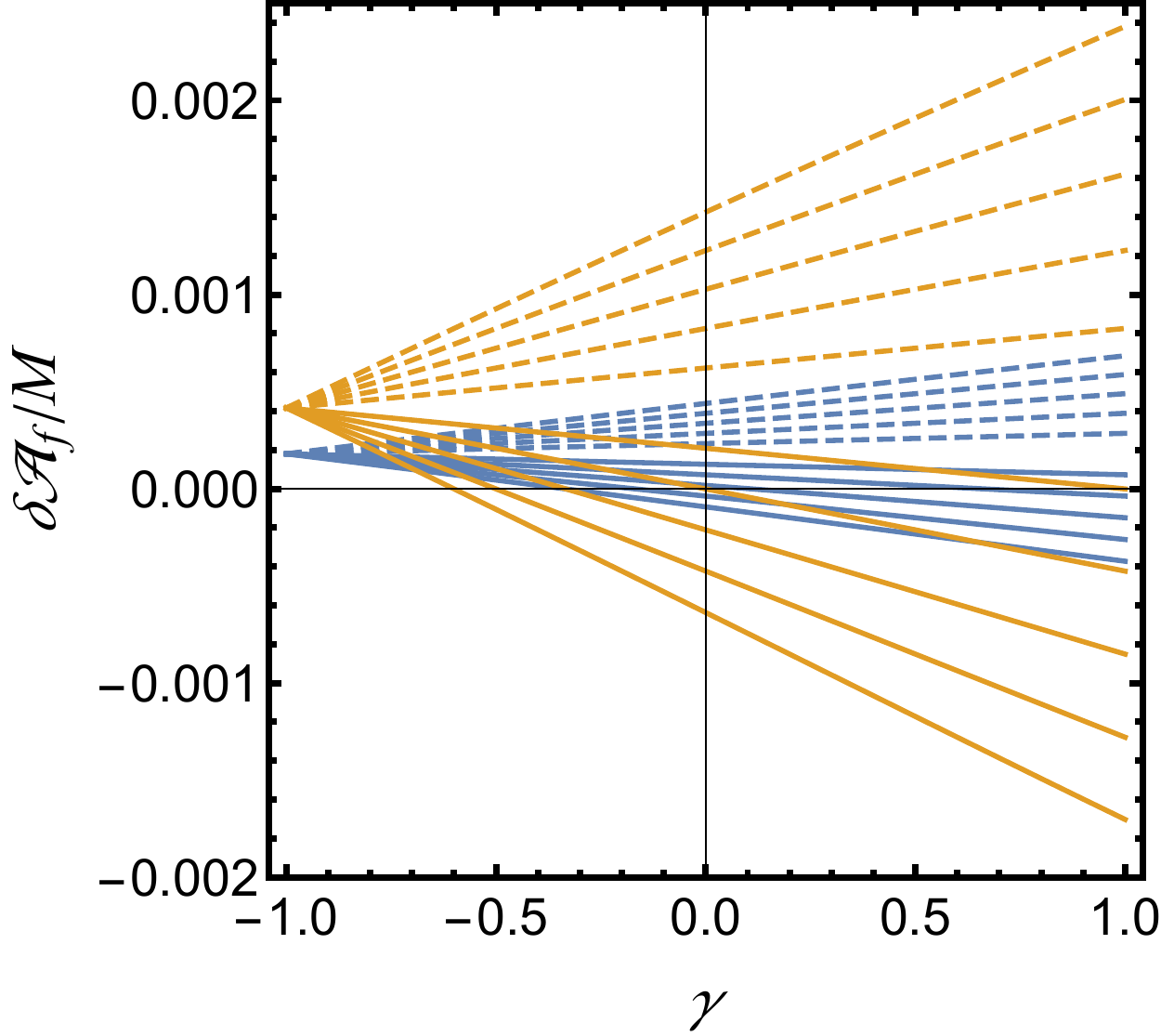} \
 \includegraphics[width=0.33\linewidth]{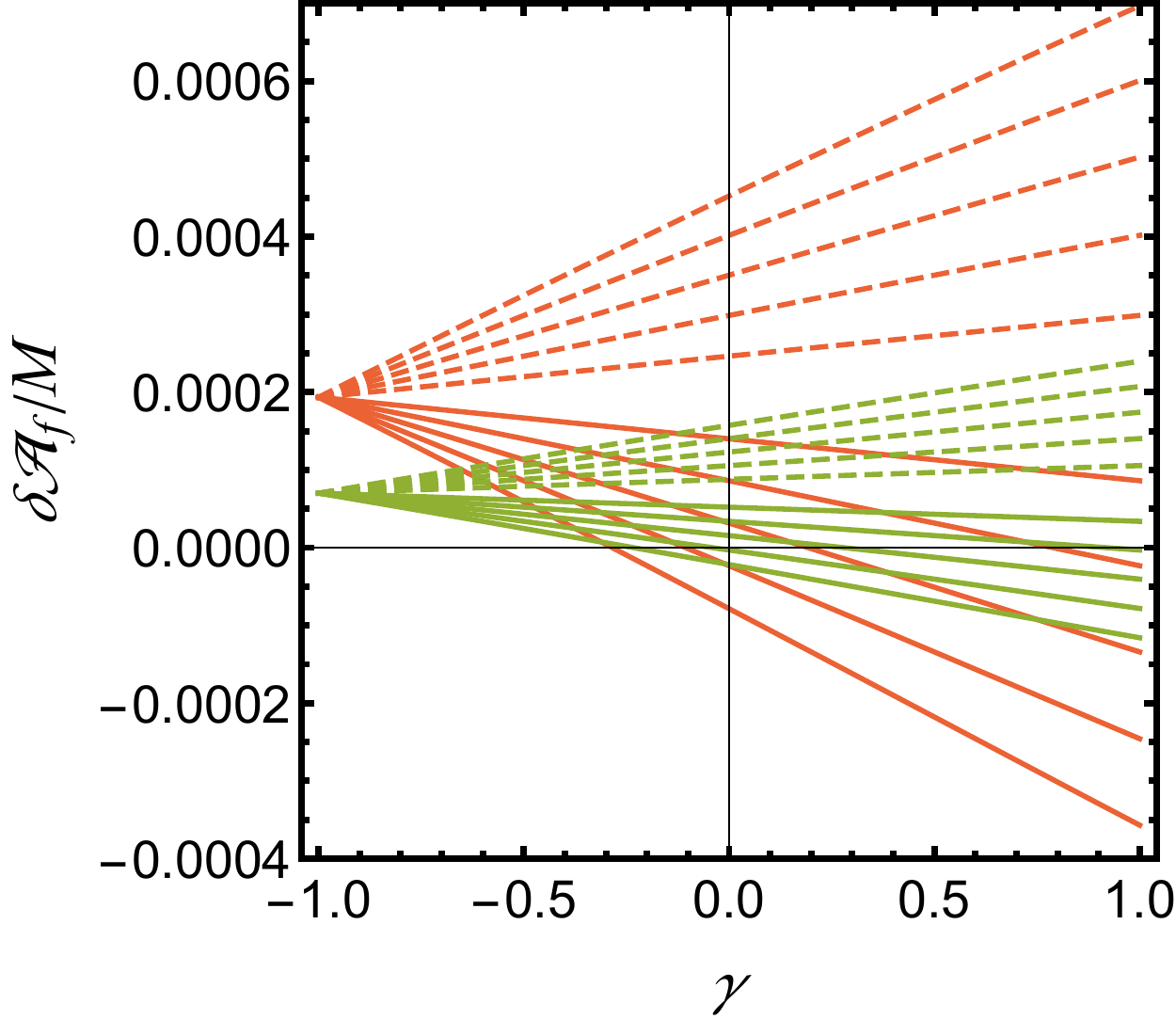}
}
\caption{In left and middle plots, the final spin's difference between two cases corresponding to whether the symmetry is taken into account or not in the EMDA (left) and KK (middle) theories. The blue lines and orange lines represent $Q= 0.05 M$ and $ Q = 0.1 M$ respectively. In right plot the green lines and red lines represent the final spin's difference between the KK and KN cases and between the EMDA  and KN cases. In right plot, the symmetry is taken into account and the charge $ Q = 0.05 M$. We set $e =0.1$. The initial spin $\chi$ of the lines in same color is 0.5, 0.4, 0.3, 0.2, 0.1, -0.1, -0.2, -0.3, -0.4, -0.5 respectively from bottom to top.}
\label{Afnu2xxx}
\end{figure}

In the previous works~\cite{Jai-akson:2017ldo, Siahaan:2019oik}, the final spin estimation of the charged BBH merger was only considered  in the equal initial spins case. Here, we will  estimate the final spin of the BBH merger with unequal initial spins, i.e. $\chi_1 =\chi, \chi_2 = \gamma \chi$, in this subsection,  and study the generic case in next subsection.
Now, the final spin estimation formula can be rewritten as:
\be
{\cal A}_f  = \frac14 ({\cal L}  +M \chi + \gamma M \chi ) \,.
\ee
For simplicity, we assume that the initial black holes have equal masses ($\nu = 1/4$), weak charges ($Q=0.05 M$ and $0.1 M$) and small  spins $|\chi_i| \le 0.5 $. From the above equation, it follows that if the initial spins of two black holes are equal and opposite, the final spin is determined totally by the angular momentum $\cal L$ of the test particle regardless of the initial spins.  The final spin increases when $\chi$ is positive and decreases when $\chi$ is negative. We plot the final spin $A_f$ against initial spin ratio $\gamma$ for different initial spins in the EM, EMDA  and KK cases in Fig.~\ref{Afnu2}.
We impose that the test particle preserves the symmetry in all cases. Note that  similar behaviors occur in the cases where the symmetry is not taken into consideration, So, we do not plot them here.
From Fig.~\ref{Afnu2}, we find that the final spin will be smaller as the BBH system carries more charges, which is same as that in the equal spin configuration. These properties exist in all cases irrespective of whether the test particle preserves the symmetry or not. We also study the difference of the final spin between two cases corresponding to whether the test particle preserves symmetry or not, and plot  the final spin's difference against $\gamma$ in the EMDA and KK cases in Fig.~\ref{Afnu2xxx}.
The difference is subtle. While the initial spins are opposite, the final spin estimated by the BKL recipe in which the symmetry is imposed is larger than that in which the symmetry is not  imposed. For positive initial spins, the final spin estimated by BKL recipe in which the symmetry is imposed can be smaller than that  in which the symmetry is not  imposed.

\subsection{Generic initial spins }

\begin{figure}[htbp]
\centerline{
 \includegraphics[width=0.33\linewidth]{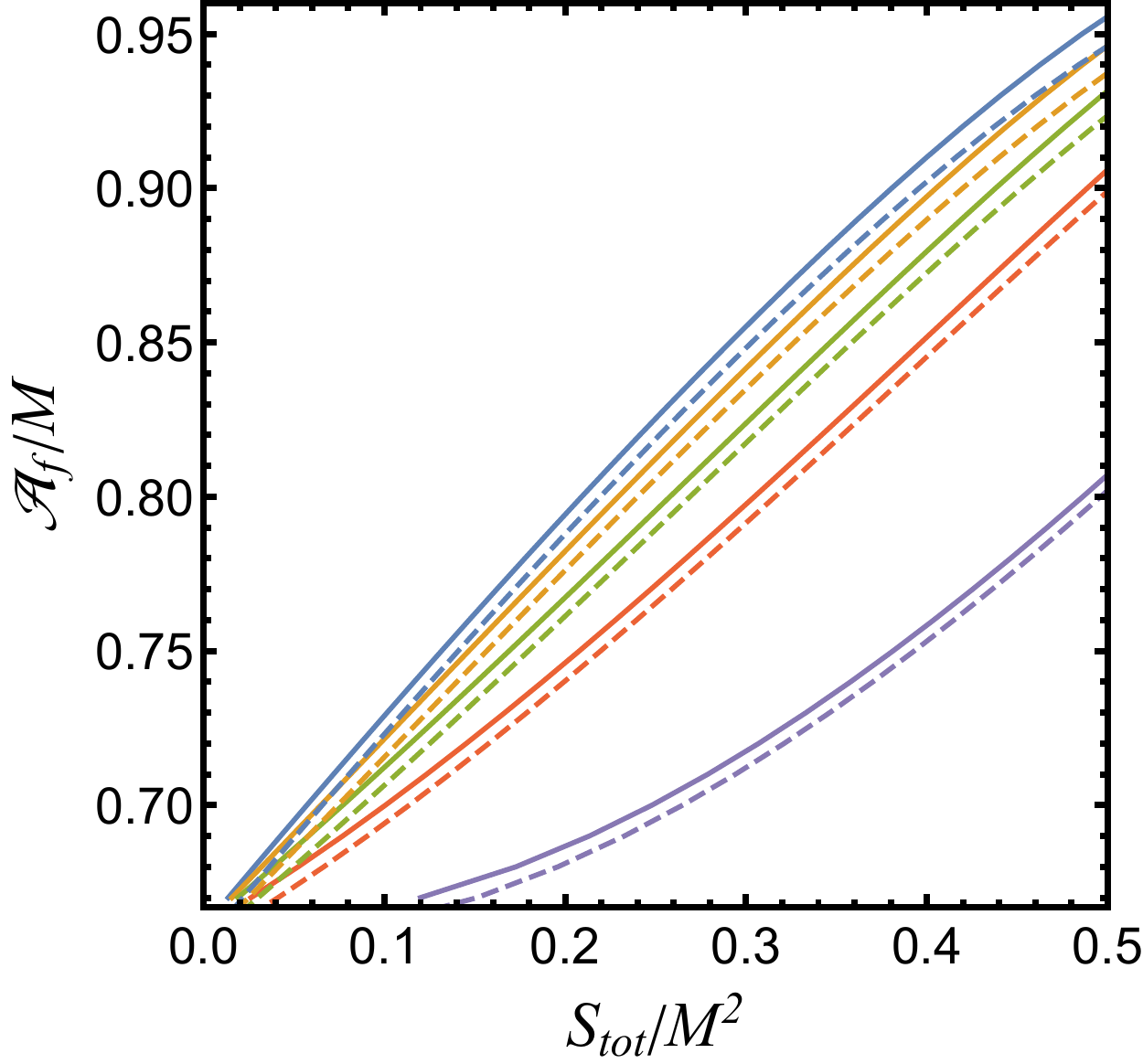} \
 \includegraphics[width=0.33\linewidth]{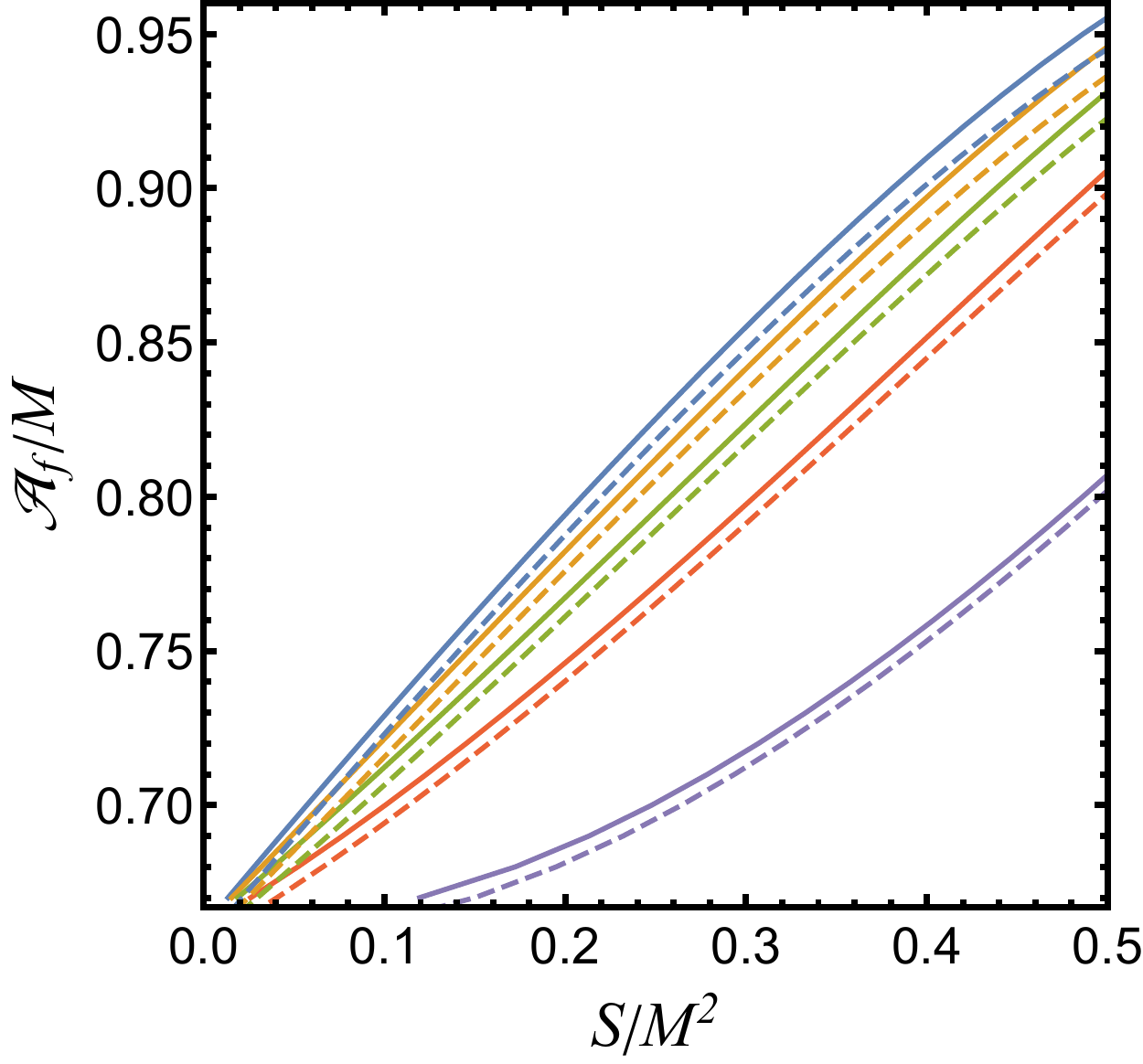} \
 \includegraphics[width=0.33\linewidth]{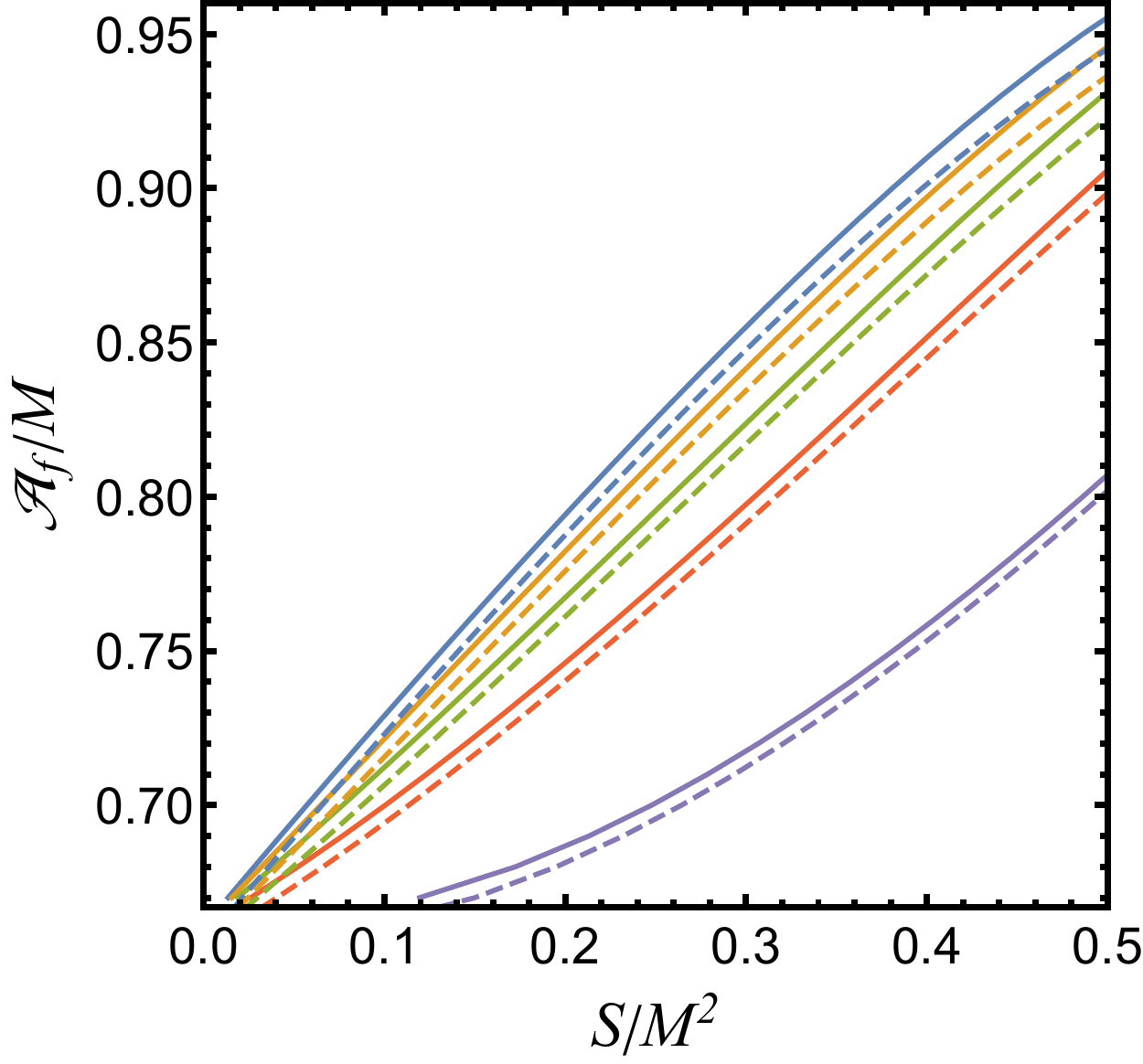}
}
\caption{The solid lines and dashed  lines  represent $Q= 0.05 M$ and $ Q = 0.1 M$ respectively. Left: KN; middle: EMDA; right: KK. We set $e =0.1$.  The inclined angle in solid lines or dashed lines  are $0, 30^\circ, 45^\circ, 60^\circ, 90^\circ$ respectively from left to right.}
\label{Afnu2xxx2}
\end{figure}

In this subsection, we consider a more generic initial spin configuration, i.e. the orbit at the ISCO can be inclined with respect to the final total angular momentum. Following Ref.~\cite{Buonanno:2007sv}, we assume the initial masses, spins and unit orbital angular momentum $(M_1, M_2, \vec{S}_1, \vec{S}_2, \hat{L}_{\textup{orb}})$ at some point of the inspiral are known. On the other hand, we assume that the magnitude of the total spin $\vec{S}_{\textup{tot}} = \vec{S}_1 + \vec{S}_2$ and angle $\vartheta_{LS}$ between total spin and unit orbital angular momentum $\hat{L}_{\textup{orb}}$ remain constant. The total angular momentum $ \vec{J}_f =M {\cal A}_f = \vec{L}_{\textup{orb}} +\vec{S}_{\textup{tot}} $, which could be written explicitly as~\cite{Buonanno:2007sv}
\bea
&&L_{\textup{orb}} \cos \vartheta +S_{\textup{tot}} \cos (\vartheta_{LS} -\vartheta) =M {\cal A}_f \,, \\
&&L_{\textup{orb}} \sin \vartheta - S_{\textup{tot}} \sin (\vartheta_{LS} -\vartheta) =0 \,,
\eea
where $L_{\textup{orb}} = |\vec{L}_{\textup{orb}}|$ and $S_{\textup{tot}} = |\vec{S}_{\textup{tot}}|$.
For simplicity, we adopt the simple fit formula given in Refs.~\cite{Buonanno:2007sv, Hughes:2002ei, Wei:2018aft}.
The orbital angular momentum of the inclined orbit is given by
\be
{\cal L}  = \frac12 (1+ \cos \vartheta) {\cal L}^{\textup{pro}} +\frac12 (1- \cos \vartheta) |{\cal L}^{\textup{ret}}| \,,
\ee
where $\vartheta$ is the inclination angle, representing the angle between final spin and the orbital angular momentum, and  ${\cal L}^{\textup{pro}}$ and  $ {\cal L}^{\textup{ret}}$ represent the angular momentum of the prograde orbit and retrograde  orbit respectively. Here we only consider the merger of a BBH with equal masses, spins and charges. The final spin can be rewritten as
\be
{\cal A}_f  =  \frac{1}{8} \left({\cal L}^{\textup{pro}} + |{\cal L}^{\textup{ret}}|  +({\cal L}^{\textup{pro}} - |{\cal L}^{\textup{ret}}|) \cos \vartheta \right ) \left(\cos \vartheta + \frac{\cos (\vartheta_{LS} -\vartheta)}{\sin (\vartheta_{LS} -\vartheta)} \sin \vartheta \right) \,.
\ee
We plot the final spin $A_f$ against the  total spin ${S}_{\textup{tot}}$ for different inclined angles $\vartheta_{LS}$ in the EM, EMDA and KK theories in Fig.~\ref{Afnu2xxx2}
by requiring the particle to preserve the symmetry.   Behaviors are similar to those in the case when the particle is not required to preserve  the symmetry. We do not plot them here.
From Fig.~\ref{Afnu2xxx2}, we find that the final spin will be smaller as the BBH system carries more charges which is again the same as the feature in equal spin configuration. Besides, we find that final spin increases as the inclined angle increases in all charge configurations.

\section{Conclusion} \label{sec5}

In this work, we study the final spin of a BBH merger in the framework of STU supergravity by using the BKL recipe.
Comparing with the previous work~\cite{Jai-akson:2017ldo}, we reconsider the contribution of the orbital angular momentum of the binary system to the final spin by requiring that the test particle preserves the scaling symmetry in Lagrangian of supergravity.
As a first step to explore whether the revised method could improve the precision of the final spin estimation, we study the final spin's difference in different initial spin configurations. In the equal initial spin configuration, we find that the difference is subtle. First, there is no difference between the two cases corresponding to  whether the symmetry is taken into account or not if the initial mass ratio is extreme. Second, for the static BBH merger, the final spin estimated by the BKL recipe in which the symmetry is imposed is always larger than that  in which the symmetry is not  imposed. The difference increases firstly and then decreases as the equal mass limit is approached. Third, there is also a critical value for the initial spins,  above which, the final spin is always smaller. The difference decreases constantly as the equal mass limit is approached.
All these features exist in the merger of a  binary STU black hole with different charge configurations (both the KK and EMDA cases). We also study the final spin difference between different charge configurations of STU supergravity with a certain number of charges. We find that the final spin differences between the cases of different charge configurations have similar features as that between two cases corresponding to whether the symmetry is imposed or not. It is worth comparing, in the future, the final spin given by the BKL recipe with numeric simulations~\cite{Hirschmann:2017psw}, and exploring if the BKL recipe could provide more accurate prediction of the final spin by requiring  the Lagrangian of the test particle preserves the  scaling symmetry in supergravity. Besides, our result  may provide a potential way to test string theory and supergravities near strong gravitational field regimes.
 We also study the final spin of a charged BBH merger in a case with unequal initial spin configuration and an even more generic spin configuration  which was not studied in previous works. We obtain results  similar to the equal initial spin case.

Finally, it is worth noting that we can also extract extra information in the merger stage by studying the null geodesic orbits for massless particle, known as light ring, in the final black hole.  For a massless particle, the geodesics is  $e^{\beta \phi} g_{\mu\nu} \dot{x}^\mu  \dot{x}^\nu =0 $ which could be obtained from Eq.~(\ref{xigauge}). It is easy to see that the conformal factor $e^{\beta \phi}$ plays no role in the corresponding calculation. So the scaling symmetry has no effect at the ringdown stage, and we do not study the light ring in this work.

\section*{Acknowledgments}

We are grateful to H. L\"u, Hao Wei, Lijing Shao and Bing Sun for useful discussions.  SLL, PXW and HWY are supported in part by NSFC grants No. 11435006, No. 11690034, No. 11775077, No. 11947216 and China Postdoctoral Science Foundation 2019M662785. WDT is supported in part by NSFC grants No. 11935009.

\appendix

\section{ Black hole solutions } \label{app1}

The KK rotating black hole solution is given by
\bea
&&ds^2_{\textup{KK}} =  -\frac{\Delta ( c dt -a \sin^2\theta d\phi)^2 }{\sqrt{H} (s^2 (r^2 +a^2) +\rho^2)}  + \sqrt{H} \Big[\frac{\rho^2}{\Delta} dr^2 +\rho^2 d\theta^2 +\frac{\sin^2\theta  (a dt - c(r^2 +a^2 )d\phi)^2 }{s^2 (r^2 +a^2) +\rho^2}  \Big]  \,,  \nn \\
&&A = \frac{2 m r  s }{\rho^2 + 2 m r s^2} (c dt  - a   \sin^2 \theta d\phi) \,,\quad \varphi =  -\frac{\sqrt{3}}{2} \ln H \,,  \quad   H = 1+\frac{2 r m s^2 }{\rho^2} \,, \label{kksol}
\eea
where $\hat{\cal A}^2 =A\,,  \hat{A}_1 = \hat{A}_2 = \hat{\cal A}^1 = 0 \,,  \varphi_1 = \varphi_2 = \varphi_3 = \varphi/\sqrt{3} \,, \psi_1 =  \psi_2 =  \psi_3 = 0 \,. $
The parameters $(m, \delta, a)$ can be written in terms of the physical quantities $M$, $Q$, and ${\cal A}=J/M$ as
\bea
m &=& \frac{3 M}{2} - \frac{1}{2}\sqrt{M^2 +8 Q^2} \,,  \nn  \\
c^2 &=& \frac{M^2 +2 Q^2 +M \sqrt{M^2 +8 Q^2}}{2 (M^2 -Q^2)} \,,   \nn    \\
a &=& \frac{\sqrt{2} M {\cal A}}{(M^2 -4 Q^2 +M\sqrt{M^2 +8 Q^2} )^{1/2}} \,,
\eea
where $Q\le M$.

The EMDA rotating black hole solution is given by
\bea
&&ds^2 = -\frac{\rho^2 - 2 m r}{\rho^2 H} \left(dt + \frac{2 m r a c^2 \sin^2 \theta d\phi}{\rho^2 - 2 m r} \right)^2 + \rho^2 H \left( \frac{dr^2}{\Delta} + d\theta^2 + \frac{\Delta \sin^2 \theta }{\rho^2 -2 m r} d\phi^2 \right) \,, \nn \\
&& A=  \frac{2 \sqrt{2} m r  s c }{\rho^2 + 2 m r s^2} ( dt - a \sin^2 \theta \ d\phi) \,, \quad \varphi = -\ln H \,, \quad \psi = \frac{2 m a \cos\theta s^2}{\rho^2} \,, \nn   \\
&& \hat{ A}_1 =  \hat{\cal A}^1  =0 \,, \quad \varphi_2 =\varphi_3 =0 \,, \quad \psi_2 =\psi_3 =0 \,, \label{emdasol}
\eea
where the parameters $(m, \delta, a)$ can be written as
\be
m = M -\frac{2 Q^2}{ M} \,, \quad a= {\cal A} \,, \quad c^2 = \frac{M^2}{M^2 -2 Q^2} \,,
\ee
where $  Q \le \frac{\sqrt{2}}{2} M $.

The KN black hole solution is given by
\bea
&&ds^2 = -\frac{\Delta_r}{\rho^2} ( dt -a \sin^2\theta d\phi)^2 +\frac{\rho^2}{\Delta_r} dr^2 +\rho^2 d\theta^2 +\frac{\sin^2\theta}{\rho^2} (a dt -(r^2 +a^2 )d\phi)^2 \,, \nn \\
&&A = \frac{4 Q r}{\rho^2} (d t  - a \sin^2 \theta d \phi ) \,, \quad  \Delta_r = r^2 -2 M r +a^2 +4 Q^2 \,.\label{knsol}
\eea

\section{ Final spin of BBH with large charges } \label{app2}

\begin{figure}[htbp]
\centerline{
 \includegraphics[width=0.45\linewidth]{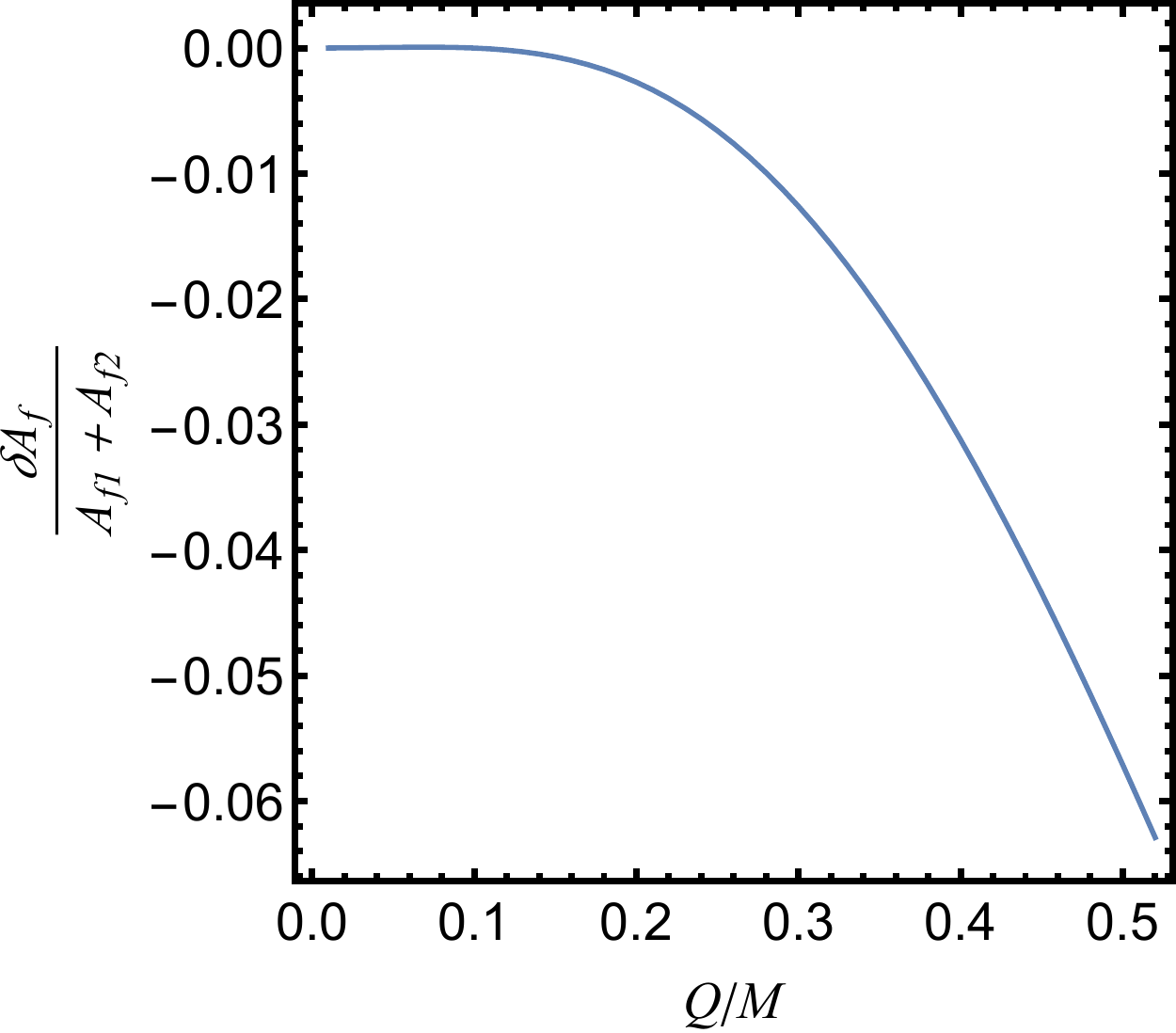} \
 \includegraphics[width=0.45\linewidth]{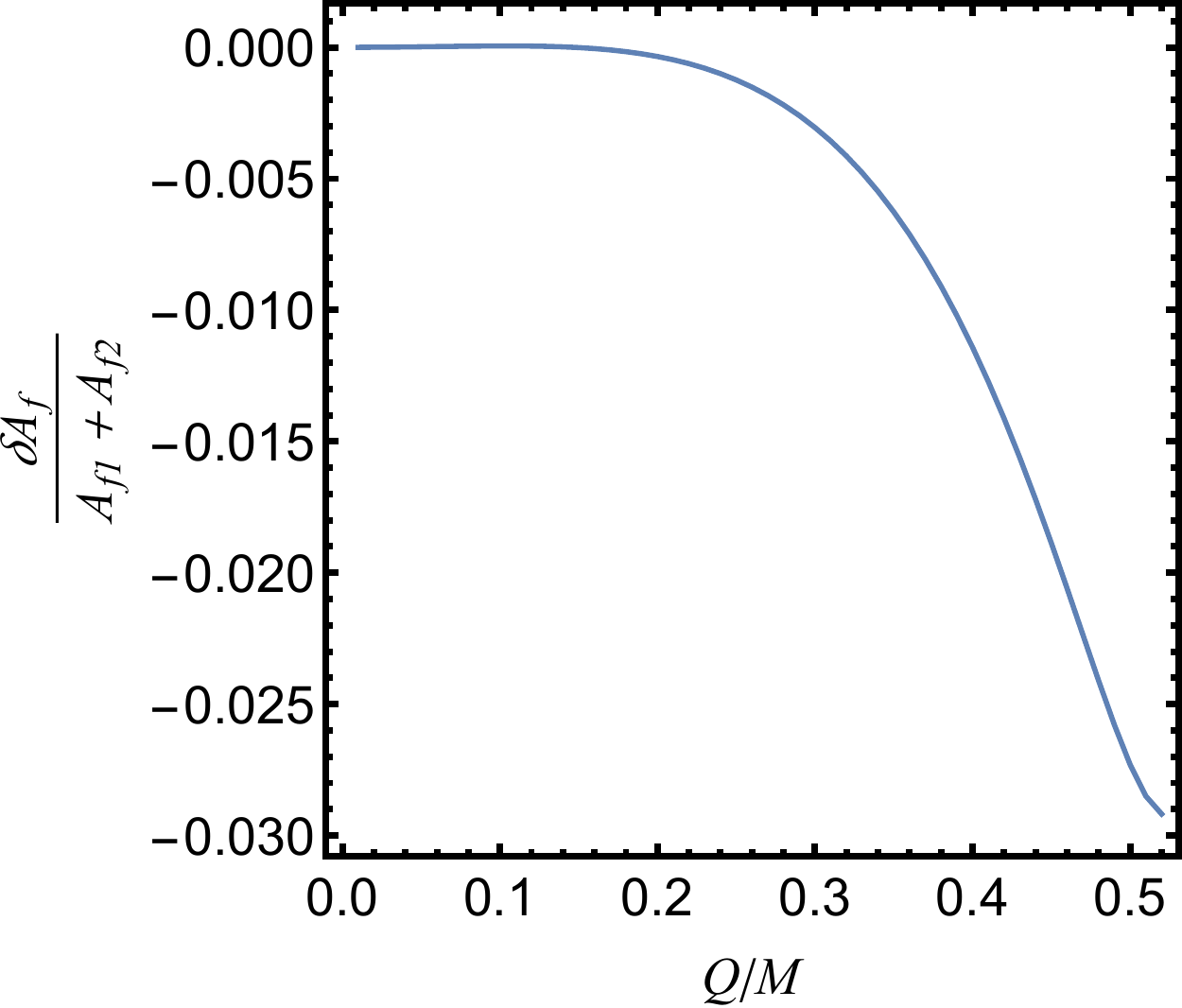}
}
\caption{ Left: KK; right: EMDA. We set $e =0.1$, initial spins $\chi = 0.1$, and consider the initial black holes have equal masses ($\nu = 1/4$).  }
\label{deltaAfQ}
\end{figure}

In this appendix, we would like to show that our approach will provide some manifestly different results if we consider the merger of binary charged black hole with  large charges.
In Figs.~\ref{deltaAfQ}, we plot the final spin's difference ${\delta {\cal A}_f}/{({\cal A}_{f1} +{\cal A}_{f2})}$ between the two cases that the scaling symmetry is taken into account or not verse the electric charge $Q$ in both KK and EMDA black holes where ${\cal A}_{f1}$ represents the symmetry is taken into account and ${\cal A}_{f2}$ represents the symmetry is not taken into account. We find the difference is about 6\% and 3\%  in KK and EMDA black holes with charge $Q \approx 0.52 M$ respectively.

\end{document}